\documentclass[10pt]{iopart}
\usepackage{epsfig}
\usepackage{iopams}
\usepackage{amsfonts}
\usepackage{amssymb}
\usepackage{mathptmx}
\usepackage{eucal}
\usepackage{bm}

\newcommand{\Eq}[1]{Eq.~(\ref{#1})}
\newcommand{\Eqs}[1]{Eqs.~(\ref{#1})}
\newcommand{\re}{\mathop{\mathrm{Re}}}

\begin{document}

\title{Relaxation of Nonequilibrium Quasiparticles in Mesoscopic Size Superconductors}

\author{K Yu Arutyunov$^{1,2}$, S A Chernyaev$^{1}$, T Karabassov$^{1}$, D S Lvov$^{3,4}$, V S Stolyarov$^{3,4}$ and A S Vasenko$^{1,5}$}
\address{$^1$ National Research University Higher School of Economics, 101000 Moscow, Russia}
\address{$^2$ P L Kapitza Institute for Physical Problems, Russian Academy of Sciences, 119334 Moscow, Russia}
\address{$^3$ Moscow Institute of Physics and Technology, 141700 Dolgoprudny, Russia}
\address{$^4$ Institute of Solid State Physics, Russian Academy of Sciences, 142432 Chernogolovka, Russia}
\address{$^5$ I E Tamm Department of Theoretical Physics, P N Lebedev Physical Institute, Russian Academy of Sciences, 119991 Moscow, Russia}

\ead{karutyunov@hse.ru}\vspace{10pt}

\begin{abstract}
Rapid development of micro- and nanofabrication methods have provoked interest and enabled experimental
studies of electronic properties of a vast class of (sub)micrometer-size solid state systems. Mesoscopic-size
hybrid structures, containing superconducting elements, have become interesting objects for basic research
studies and various applications, ranging from medical and astrophysical sensors to quantum computing.
One of the most important aspects of physics, governing the behavior of such systems, is the finite
concentration of nonequilibrium quasiparticles, present in a superconductor even well below the temperature
of superconducting transition. Those nonequilibrium excitations might limit the performance of a variety of
superconducting devices, like superconducting qubits, single-electron turnstiles and microrefrigerators. On
the contrary, in some applications, like detectors of electromagnetic radiation, the nonequilibrium state is
essential for their operation. It is therefore of vital importance to study the mechanisms of nonequilibrium
quasiparticle relaxation in superconductors of mesoscopic dimensions, where the whole structure can be
considered as an `interface'. At early stages of research the problem was mostly studied in relatively
massive systems and at high temperatures close to the critical temperature of a superconductor. We review
the recent progress in studies of nonequilibrium quasiparticle relaxation in superconductors including the
low temperature limit. We also discuss the open physical questions and perspectives of development in the
field.
\end{abstract}

\vspace{2pc}

\maketitle
\ioptwocol
\submitto{\JPCM}

%%%%%%%%%%%%%%%%%%%%%%%%%%%%%%%%%%%%%%%%%%%%%%%%%%%%%%%%%%%%%%%%%%%%%%%%%%%%
%%%%%%%%%%%%%%%%%%%%%%%%%%%%%%%%%%%%%%%%%%%%%%%%%%%%%%%%%%%%%%%%%%%%%%%%%%%%

\section{Introduction}\label{SecIntro}

%%%%%%%%%%%%%%%%%%%%%%%%%%%%%%%%%%%%%%%%%%%%%%%%%%%%%%%%%%%%%%%%%%%%%%%%%%%%
%%%%%%%%%%%%%%%%%%%%%%%%%%%%%%%%%%%%%%%%%%%%%%%%%%%%%%%%%%%%%%%%%%%%%%%%%%%%

Mesoscopic-size superconducting hybrid structures have become increasingly important devices in applications ranging
from medical and astrophysical sensors to quantum computing due to their minimal energy dissipation at low temperatures.
The conversion of electric current at an interface between different materials is a common process in any hybrid structure.
Of particular interest are boundaries with a superconductor where electric current converts from
single electrons to Cooper pairs \cite{Tinkham_book}. At nanoscales the whole system might behave as an `interface' if
the dimension(s) are comparable to the characteristic relaxation length.

In a superconductor at a finite temperature there are always nonpaired electrons called
equilibrium quasiparticles. In the presence of additional disturbance their concentration can be increased by nonequilibrium
quasiparticles. There can be deviations from equilibrium (nonequilibrium modes) of charge, energy
and/ or spin degrees of freedom.
Rapid development of micro- and nano-fabrication methods facilitated the fabrication of devices and circuits with dimensions
of the order of relaxation lengths of these nonequilibrium modes. The nonequilibrium quasiparticles might limit the performance of a variety of nanoscale superconducting devices with dimensions comparable to corresponding relaxation scales, such as refrigerators based on normal metal (N) - insulator (I) - superconductor (S) junctions \cite{Muhonen2012, Giazotto2006, Virtanen2007, Courtois2014, Rajauria2008, Vasenko2010}, NIS refrigerators with a ferromagnetic (F) interlayer \cite{Ozaeta2012, Kawabata2013, Kawabata2015}, superconducting resonators \cite{Nsanzineza2014, Sandberg2009, Harvey2008, Levenson-Falk2012}, superconducting qubits \cite{Makhlin2001, Wendin2007, Paauw2009, Martinis2009, Paik2011, Corcoles2011, deVisser2011, Wenner2013, Wang2014, Vool2014}, single-electron hybrid turnstiles \cite{Knowles2012, Taupin2016}, SFS $\pi$-junctions \cite{Golubov_rev, Ryazanov2001, Vasenko2008, Vasenko2011}, nanorings \cite{Arutyunov2004, Vodolazov2006, Arutyunov2012} and many other devices. The effect has been
notoriously called quasiparticle poisoning. On the contrary, in some applications, like various types of photon detectors and bolometers \cite{Golubov1993, Golubov1994, Goltsman2001, Richards1994, Kuzmin2004}, the nonequilibrium state is essential for their operation.

The problem of quasiparticle removal in superconducting devices can be partly solved by introducing the so called quasiparticle traps away from
the junction region, either by using normal-metal layers covering the superconducting electrode \cite{Goldie1990, Pekola2000, Ullom2000, Rajauria2012} or the local energy gap suppression by an external magnetic field \cite{Peltonen2011, Nakamura2017}. Quasiparticles are then trapped by the region with no energy gap (or suppressed gap). Other possibility is an alternative device design immune to quasiparticle overheating \cite{vanZanten2016}.

The phenomena of relaxation of nonequilibrium quasiparticles attracted attention in the mid-1970s resulting in an
impressive number of papers (for references see section \ref{QPrelax}). Those early
experiments were mainly performed on sandwich-type flat structures not adequate for spatially resolved studies. Agreement between the experiment and theory was established reliably, mainly in the high temperature limit $T \rightarrow T_c$. The understanding of the opposite limit $T \ll T_c$
is still far from being satisfactory. In this topical review we focus on our works, where for the first time we have measured relaxation lengths for the charge and energy nonequilibrium modes on the same hybrid microstructure at ultra-low temperatures. We compare these results with new recent results of nonlocal measurements. We also review other works related to the quasiparticle relaxation modes.

It should be mentioned that recently it has been published many papers on spin accumulation and spin relaxation in superconductors with a spin-splitting field (for comprehensive reviews see \cite{Beckmann_rev, Quay_rev, Bergeret_rev}). We will not discuss the deviations from equilibrium of the spin degrees of freedom in this topical review and will focus on present understanding of charge and energy modes relaxation in superconductors without spin-splitting.

The review is organized as follows: in the next section (Sec.~\ref{SecNoneqQP}) we describe the energy and charge nonequilibrium modes in superconductors and the ways of their excitation. In section \ref{QPrelax} we introduce a typical time hierarchy, that describes these modes relaxation into a ground state. Then in section \ref{f_noneq} we introduce the nonequilibrium distribution functions and basic equations for electric, energy and heat currents. In section \ref{ph_model} we discuss experiments of other authors and the existing phenomenological models. In section \ref{setup} we introduce the multiterminal nanostructures we have used to measure the relaxation times of charge and energy nonequilibrium modes (on the same sample). We summarize the definitions of various
parameters of the dimensionality `temperature' used in this review in section \ref{temperatures} and discuss the nonequilibrium quasiparticle injection in section \ref{injection}. We present our principle results concerning the energy (section \ref{energy}) and charge (section \ref{charge}) imbalance relaxation lengths. In section \ref{nonlocal} we present new experimental results on nonlocal supercurrent measurements for charge imbalance length determination. In section \ref{cooling} we review recent results on nonequilibrium electron cooling in NIS refrigerators and discuss the problems of quasiparticle poisoning and evacuation. Finally, in section \ref{summary} we give a summary and an outlook on possible future developments in the field.

\begin{figure}[b]\label{QP}
\begin{center}
\epsfxsize=6cm\epsffile{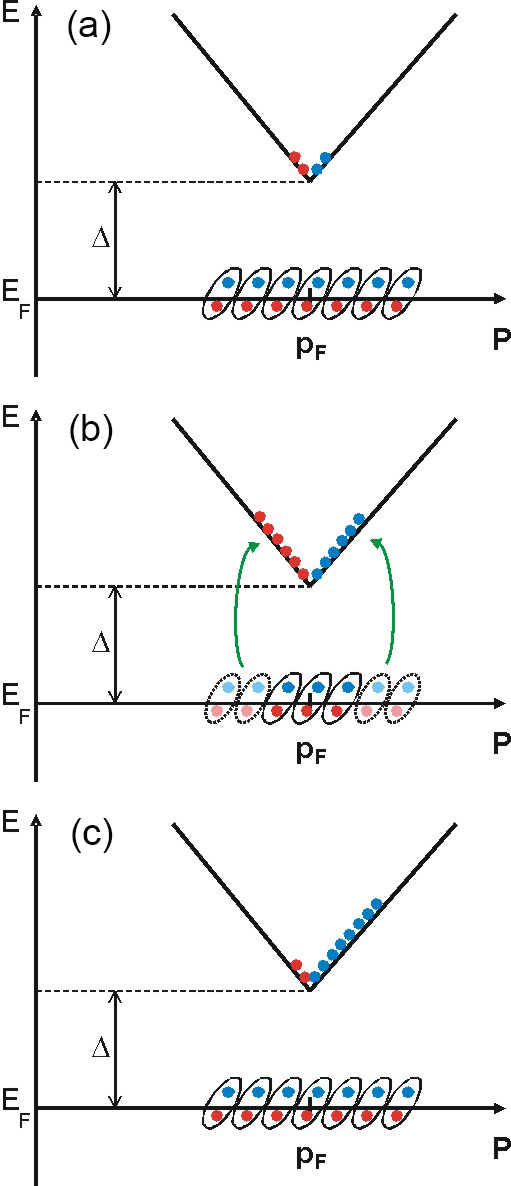}
\end{center}
\caption{Schematic representation of an  energy spectrum of a superconductor: excited states are separated from the ground state by the energy gap $\Delta$ at a Fermi level $E_{F}$. (a) Equilibrium quasi-particle excitations which are present at any finite temperature. (b) Nonequilibrium quasiparticles which arise due to the Cooper pair breaking and symmetrically occupy excited states with respect to a Fermi momentum $p_F$. (c) The charge imbalance - an asymmetrical occupation of the excitation spectrum.}
\end{figure}
%

%%%%%%%%%%%%%%%%%%%%%%%%%%%%%%%%%%%%%%%%%%%%%%%%%%%%%%%%%%%%%%%%%%%%%%%%%%%%
%%%%%%%%%%%%%%%%%%%%%%%%%%%%%%%%%%%%%%%%%%%%%%%%%%%%%%%%%%%%%%%%%%%%%%%%%%%%

\section{Nonequilibrium quasiparticles in superconductors}\label{SecNoneqQP}

%%%%%%%%%%%%%%%%%%%%%%%%%%%%%%%%%%%%%%%%%%%%%%%%%%%%%%%%%%%%%%%%%%%%%%%%%%%%
%%%%%%%%%%%%%%%%%%%%%%%%%%%%%%%%%%%%%%%%%%%%%%%%%%%%%%%%%%%%%%%%%%%%%%%%%%%%

%
\begin{figure*}[h!]\label{Kaplan_times}
\begin{center}
\epsfxsize=16cm\epsffile{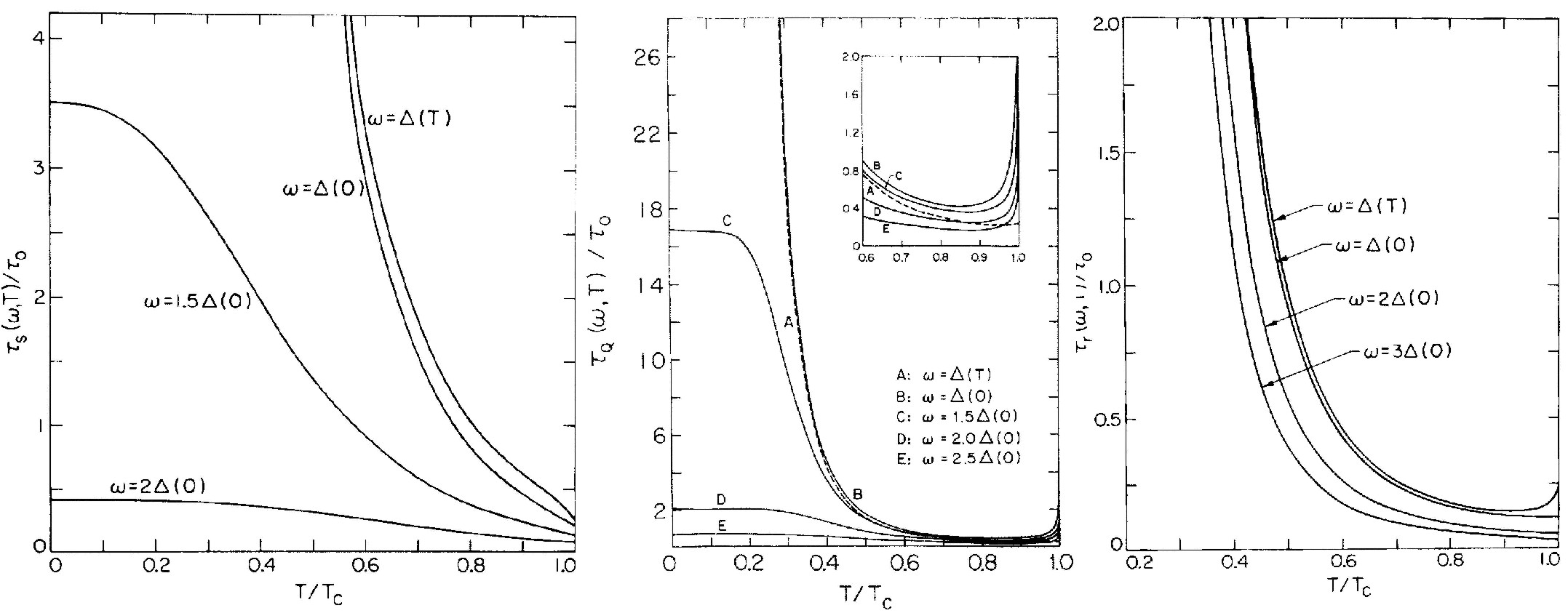}
\end{center}
\caption{Calculated energy dependencies of characteristic relaxation times $\tau_s$, $\tau_Q$ and $\tau_r$ expressed in units of the electron-phonon scattering time $\tau_0$ \cite{Kaplan1976}.}
\end{figure*}

In Fig.~\ref{QP} we schematically represent an energy spectrum of a superconductor: excited states, separated from the ground state by the energy gap $\Delta$ and equilibrium charge carriers (Cooper pairs) at a Fermi level $E_{F}$. A certain number of unpaired electrons (equilibrium quasiparticles) are always present in a superconductor at any finite temperature $T < T_c$ due to a thermal activation [Fig.~\ref{QP}(a)]. These equilibrium excitations does not contribute to the electronic transport and, consequently the resistance of the superconductor (dc biased) equals to zero. The occupancy of excited states can be expanded at the same arbitrary temperature $T < T_c$ by creating nonequilibrium quasiparticles. If such quasiparticles are created due to processes connected with the Cooper pair breaking (for example, irradiation by photon energies larger than the energy gap, $\hbar \omega > \Delta$), they symmetrically occupy excited states with respect to a Fermi momentum $p_F$ [Fig.~\ref{QP}(b)]. If transport measurements are conducted fast enough than the finite voltage drop (i.e. the resistance) in the superconductor can be registered at a time scale smaller than corresponding relaxation times. However, there appears to be an even more nontrivial deviation from the equilibrium, the so called charge imbalance, which is an asymmetrical occupation of the excitation spectrum [Fig.~\ref{QP}(c)]. It can be created, for example, by electron injection from a normal metal into a superconductor. Depending on a polarity of an applied voltage, electron-like quasiparticles with the momentums $p > p_F$ [as shown in Fig.~\ref{QP}(c)] as well as hole-like quasiparticles with the momentums $p < p_F$  can be created. Historically, several notations were introduced for both excitation types [Fig.~\ref{QP}(b) and Fig.~\ref{QP}(c)]: energy and charge imbalance or longitudinal and transverse modes of quasi-particle excitations \cite{Schmid1975, Arutyunov1996}. In what follows we will use both type of notations.

%%%%%%%%%%%%%%%%%%%%%%%%%%%%%%%%%%%%%%%%%%%%%%%%%%%%%%%%%%%%%%%%%%%%%%%%%%%%
%%%%%%%%%%%%%%%%%%%%%%%%%%%%%%%%%%%%%%%%%%%%%%%%%%%%%%%%%%%%%%%%%%%%%%%%%%%%

\section{Quasiparticle relaxation times}\label{QPrelax}

%%%%%%%%%%%%%%%%%%%%%%%%%%%%%%%%%%%%%%%%%%%%%%%%%%%%%%%%%%%%%%%%%%%%%%%%%%%%
%%%%%%%%%%%%%%%%%%%%%%%%%%%%%%%%%%%%%%%%%%%%%%%%%%%%%%%%%%%%%%%%%%%%%%%%%%%%

%
\begin{figure}[t!]\label{Clarke-Tinkham}
\begin{center}
\epsfxsize=6cm\epsffile{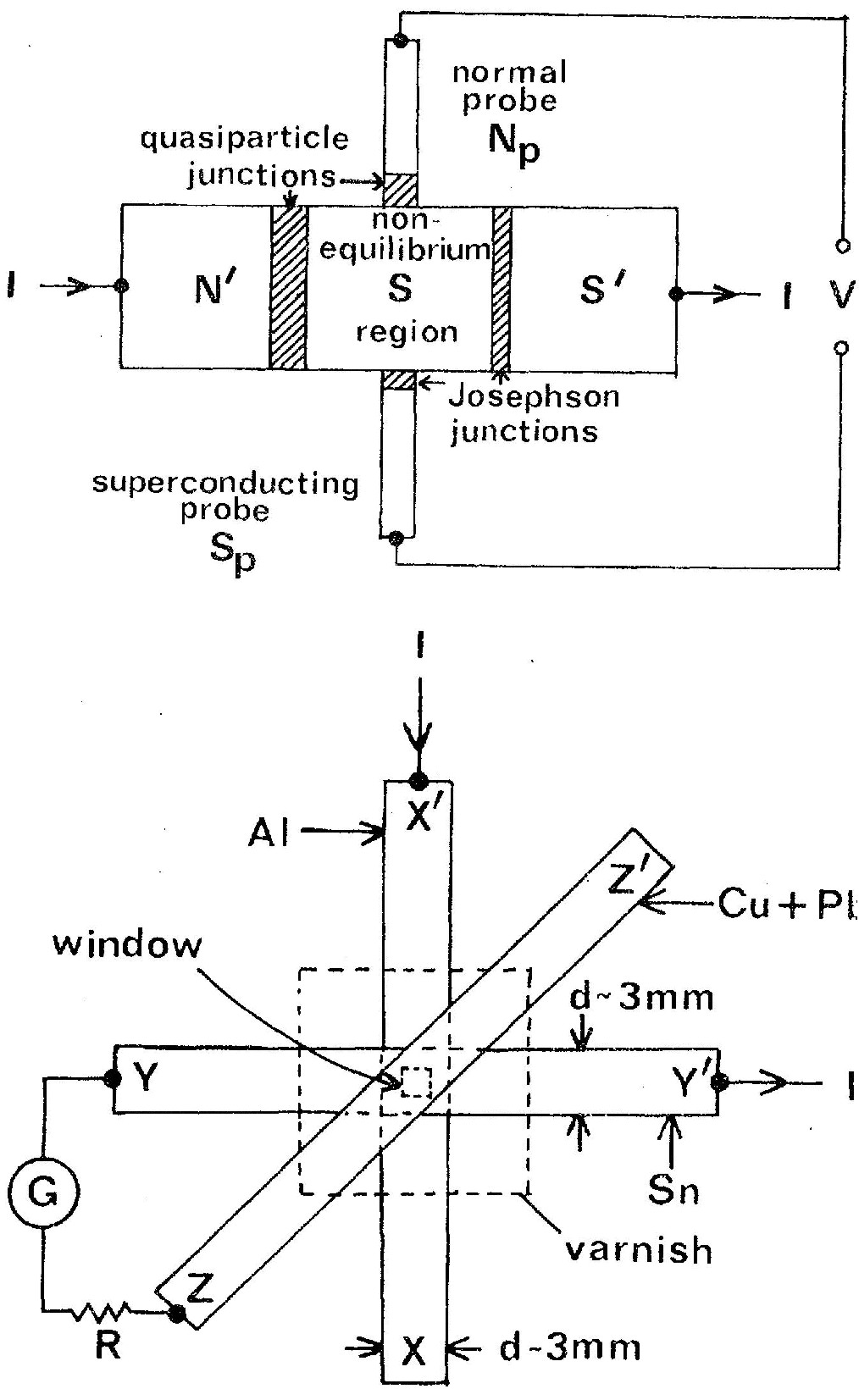}
\end{center}
\caption{The design of the experiment and the sample in \cite{Tinkham1972, Clarke1972}.}
\end{figure}

The first systematic research in the field of nonequilibrium superconductivity began in the early seventies. A highly detailed state of this matter can be found in profound reviews \cite{Chang1978}, \cite{Clarke1981} and \cite{Kopnin2001}. Despite the significant progress in understanding of the physics of processes, that was achieved in the following decades \cite{Galitski1973, GalitskiUFN, Elesin1981, Mitsen1981, Golovashkin1981}, a number of questions are still open. The main subject of this topical review is the studies of current states of a superconductor. We will concentrate on questions connected with charge imbalance, which is typical for a current injection of nonequilibrium excitations (so called transverse nonequilibrium mode, see section \ref{f_noneq}). Nevertheless, charge imbalance is always accompanied by the excitation of energy imbalance (the longitudinal mode, see section \ref{f_noneq}), which, for example, is related to the electron cooling phenomenon. Both excitations will be discussed in this review.

Postulating the existence of a definite energy spectrum of the superconductor in Fig.~\ref{QP}, it can be assumed from general considerations that some typical time hierarchy, that describes the nonequilibrium modes relaxation into a ground state, should exist. Apparently, the smallest scale can be the time of an electron-electron interaction $\tau_{e-e}$. Electrons are not thermalized at shorter times and, consequently cannot be described by the definite distribution function. A nonequilibrium population of the energy spectrum which corresponds both to a longitudinal [Fig.~\ref{QP}(b)] and a transverse [Fig.~\ref{QP}(c)] mode may take place after the thermalization. Obviously, equilibrium Cooper pairs are formed from electrons which moments are equal in absolute values but are opposite in the sign (with respect to the Fermi momentum $p_F$). They cannot be formed directly from a type of spectrum illustrated in Fig.~\ref{QP}(c): first of all the numbers of electron-like and hole-like excitations should be equal at typical relaxation times of a charge imbalance $\tau_Q$. A parallel process is an inelastic quasiparticle scattering with a phonon emission and an absorption which is defined by the time $\tau_s$. And finally, an inelastic recombination of two quasiparticles which leads to the forming of an equilibrium Cooper pair is described by the typical time $\tau_r$, see Fig.~\ref{Kaplan_times} \cite{Kaplan1976}. In the most general case all these three processes are inelastic and do require a presence of a subsystem which allows the energy exchange, for example, phonons. It is remarkable, that in case of a gap anisotropy and/ or a presence of a finite current and/ or magnetic impurities the relaxation of a charge imbalance may occur due to the elastic processes \cite{Tinkham1972}. Obviously, with a temperature drop this relaxation channel should become the dominant one.

%%%%%%%%%%%%%%%%%%%%%%%%%%%%%%%%%%%%%%%%%%%%%%%%%%%%%%%%%%%%%%%%%%%%%%%%%%%%
%%%%%%%%%%%%%%%%%%%%%%%%%%%%%%%%%%%%%%%%%%%%%%%%%%%%%%%%%%%%%%%%%%%%%%%%%%%%

\section{Basic equations}\label{f_noneq}

%%%%%%%%%%%%%%%%%%%%%%%%%%%%%%%%%%%%%%%%%%%%%%%%%%%%%%%%%%%%%%%%%%%%%%%%%%%%
%%%%%%%%%%%%%%%%%%%%%%%%%%%%%%%%%%%%%%%%%%%%%%%%%%%%%%%%%%%%%%%%%%%%%%%%%%%%

%
\begin{figure}[t!]\label{Clarke-Vex}
\begin{center}
\epsfxsize=6cm\epsffile{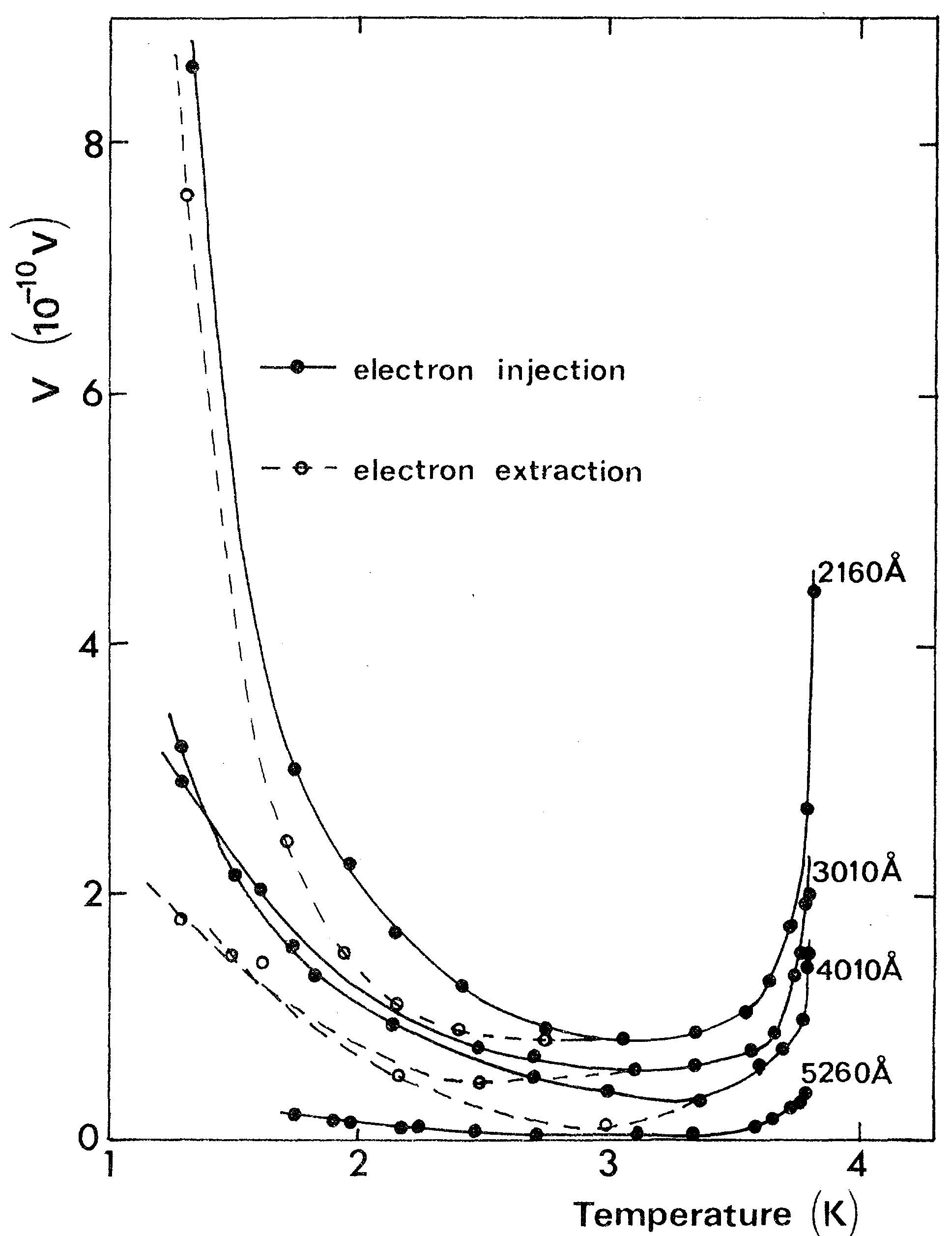}
\end{center}
\caption{A temperature dependence of the excess voltage in the nonequilibrium superconductor \cite{Clarke1972}.}
\end{figure}

Nonequilibrium distribution functions $f_L$ and $f_T$ for longitudinal and transverse modes, respectively, can be defined within the framework of a Keldysh formalism \cite{Belzig1999, Eschrig2000, Bezuglyi2005, Bezuglyi2006, Bezuglyi2017, Bezuglyi2007}. The matrix Keldysh-Green's function (which is 2$\times$2 matrix in Nambu space) is given by the following expression,
\begin{equation}\label{Keld}
\hat{G}^K = \hat{g}^R \hat{f} - \hat{f} \hat{g}^A,
\end{equation}
where $\hat{g}^R$ and $\hat{g}^A$ are, correspondingly, the retarded and advanced Green's functions in Nambu space, and $\hat{f}$ is the matrix distribution function
having only diagonal elements
\begin{equation}\label{f}
\hat{f} = f_L + \sigma_z f_T,
\end{equation}
where $f_{L (T)}$ are odd (even) components (scalars) with respect to the Fermi surface, and $\sigma_z$ is the Pauli matrix. Collecting $\hat{g}^R$, $\hat{g}^A$ and $\hat{G}^K$ into the symbolic 4$\times$4 matrix in Keldysh-Nambu space, one can write the diffusive equations of nonequilibrium superconductivity (see \cite{Belzig1999} for the comprehensive review). Obtaining $f_{L,T}$ from these equations is the central question of the nonequilibrium superconductivity. We note that the normal distribution function in such representation is expressed as
\begin{equation}
f = \frac{1}{2} \left ( 1 - f_T -f_L \right ).\label{fnorm}
\end{equation}

Matrix distribution function can be expressed through the electron and hole population numbers, $n^e$ and $n^h$, respectively,
\begin{eqnarray}
\hat{f} = 1 - 2\left(\begin{array}{cc} n^e &
0 \\
0 & n^h
\end{array}\right), \label{neh1}
\\
f_L = 1- n^e - n^h,
\quad f_T = n^h - n^e.\label{neh2}
\end{eqnarray}
In thermal equilibrium, for instance in the reservoirs at voltage $V$, it can be expressed by
the Fermi functions $f_F(E) = [1 + \exp(E/k_B T)]^{-1}$,
\begin{eqnarray}
\hat{f}^{eq} = \left(\begin{array}{cc} 1 - 2 f_F(E+eV) &
0 \\
0 & 2f_F(-E+eV) - 1
\end{array}\right) \nonumber
\\
= \left(\begin{array}{cc} \tanh\frac{E+eV}{2k_B T} &
0 \\
0 & \tanh\frac{E-eV}{2k_B T}
\end{array}\right),
\end{eqnarray}
where $e$ is the electron charge, $E$ is the quasiparticle energy, counted from the Fermi energy $E_F$, $T$ is the temperature, and $k_B$ is the Boltzmann constant. Equilibrium values of $f_{L (T)}$ can be expressed as,
\begin{eqnarray}
f_{L}^{eq} = \frac{1}{2} \left( \tanh\frac{E+eV}{2 k_B T} + \tanh\frac{E-eV}{2 k_B T} \right),\label{f_L}
\\
f_{T}^{eq} = \frac{1}{2} \left( \tanh\frac{E+eV}{2 k_B T} - \tanh\frac{E-eV}{2 k_B T} \right).\label{f_T}
\end{eqnarray}

From \Eqs{neh1},\eref{neh2} it can be seen a clear physical explanation for $f_L$ and $f_T$ distribution functions.
The deviations of $f_L$ from equilibrium produces more (or
fewer) quasiparticles equally on both holelike and electronlike
branches of the quasiparticle spectrum of the superconductor, while the deviation of $f_T$ from equilibrium
produces more electrons than holes or vice versa (therefore this class of disequilibrium it is often called
as ``charge'' or ``branch'' imbalance). From the last equation, \Eq{f_T}, one can see that the equilibrium value of the transverse mode is zero at zero voltage bias.

In this review we will mostly consider the problem of nonequilibrium quasiparticle injection from a normal metal lead into a superconductor in normal metal - insulator - superconductor (NIS) tunnel junctions. The flow of electric current in NIS tunnel junctions is accompanied by the
heat transfer from the normal metal into the superconductor. The electric current within the Keldysh formalism is given by the following equation \cite{Belzig1999},
\begin{equation}
I = \frac{g_N}{e} \int_0^\infty \mathcal{D}_T \nabla f_T \; dE, \label{Keld_I}
\end{equation}
where $g_N$ is the normal conductance of the normal metal lead per unit length, and $\mathcal{D}_T = \mathrm{Tr} (1 - \sigma_z \hat{g}^R \sigma_z \hat{g}^A)$ is a dimensionless diffusion coefficient. The energy current within the Keldysh formalism is defined as \cite{Belzig1999},
\begin{equation}
Q = \frac{g_N}{e^2} \int_0^\infty E \mathcal{D}_L \nabla f_L \; dE, \label{Keld_Q}
\end{equation}
where $\mathcal{D}_L = \mathrm{Tr} (1 - \hat{g}^R \hat{g}^A)$. The spectral electric current is proportional to the transverse mode gradient (which is created, for example, by the applied voltage), while the spectral energy current is proportional to the gradient of the longitudinal mode. The heat current out of the normal metal lead in an NIS junction (the cooling power) is given by \cite{Vasenko2010},
\begin{equation}
P = - IV - Q.
\end{equation}
It is due to selective tunneling of high-energy quasiparticles out of the normal
metal which is induced by the superconducting energy gap. The heat $P$ taken
from the N electrode is then released in the superconductor electrode, thus the full heat production in both electrodes
is equal to the Joule heating, $IV$.

We mention here that the current transport in NIS junctions is governed not only by single-particle tunneling but also by two-particle
(Andreev) tunneling \cite{Hekking1993, Volkov1993}. In this topical review we do not discuss the questions of nonequilibrium current transport in junctions and weak links through the Andreev levels. The reader can see, for example, the following artiles \cite{Bretheau2013N, Bretheau2013, Avriller2015, Olivares2014, Kalenkov2007, Golubev2007, Golubev2009, Golubev2009(2), Galaktionov2017} and references therein.

%%%%%%%%%%%%%%%%%%%%%%%%%%%%%%%%%%%%%%%%%%%%%%%%%%%%%%%%%%%%%%%%%%%%%%%%%%%%
%%%%%%%%%%%%%%%%%%%%%%%%%%%%%%%%%%%%%%%%%%%%%%%%%%%%%%%%%%%%%%%%%%%%%%%%%%%%

\section{Early experiments and phenomenological models}\label{ph_model}

%%%%%%%%%%%%%%%%%%%%%%%%%%%%%%%%%%%%%%%%%%%%%%%%%%%%%%%%%%%%%%%%%%%%%%%%%%%%
%%%%%%%%%%%%%%%%%%%%%%%%%%%%%%%%%%%%%%%%%%%%%%%%%%%%%%%%%%%%%%%%%%%%%%%%%%%%

%
\begin{figure}[t!]\label{Yagi}
\begin{center}
\epsfxsize=6cm\epsffile{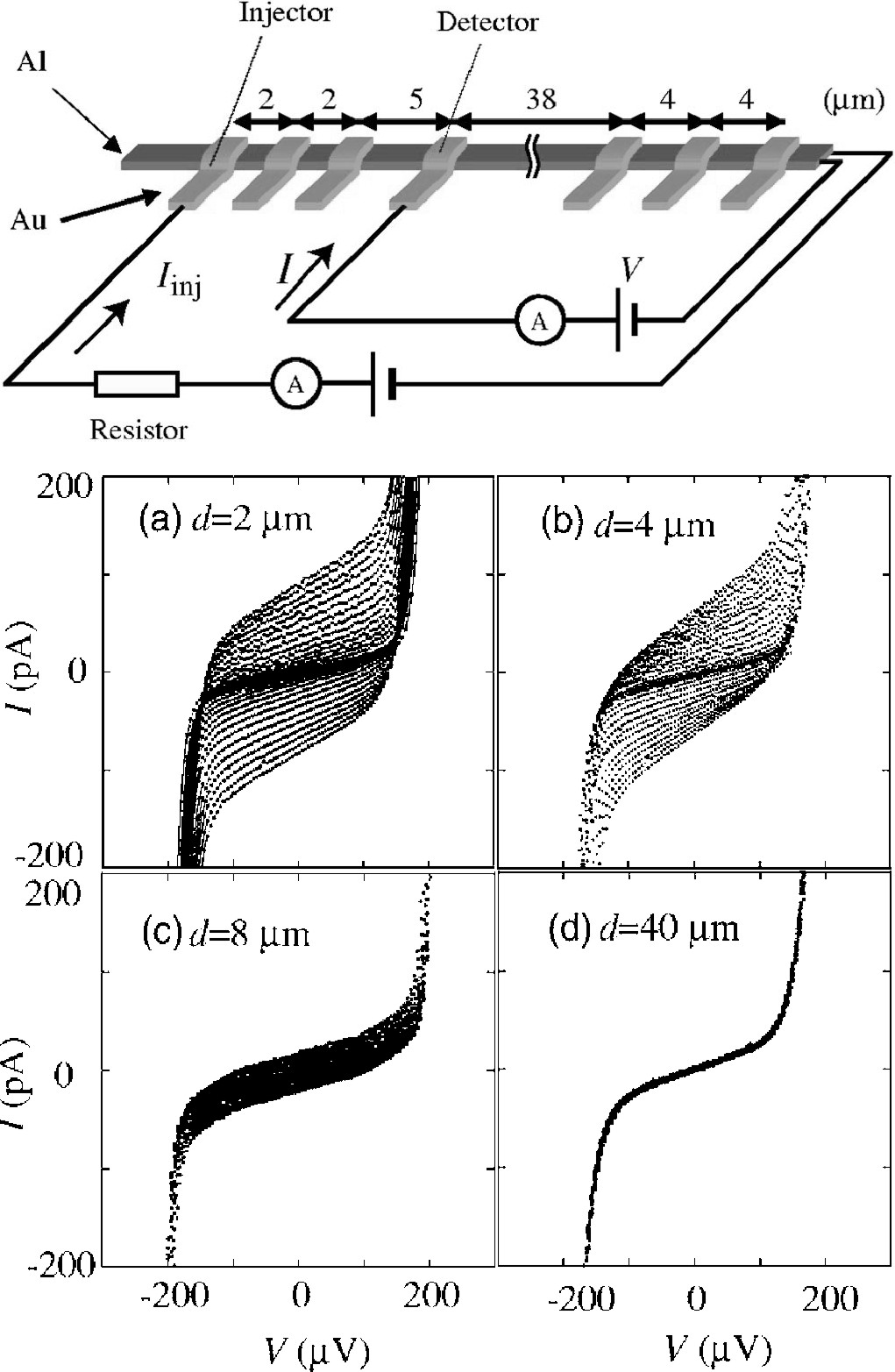}
\end{center}
\caption{An experimental design and current-voltage characteristics of the NIS detectors placed at various distances from the NIS injector \cite{Yagi2006}.}
\end{figure}

First experimental works on charge imbalance in superconductors began to appear in the early seventies \cite{Tinkham1972, Clarke1972}. The experimental design is illustrated in Fig.~\ref{Clarke-Tinkham}: nonequilibrium quasiparticles are injected from a normal metal (N) through a tunnel barrier (I) into a superconductor (S) where a potential difference is measured either by a superconducting electrode (S$_\mathrm{p}$) through a Josephson contact or by a normal electrode (N$_\mathrm{p}$) through a thin tunnel barrier.

A finite magnitude of the excess voltage [see Fig.~\ref{Clarke-Vex}] detected in these experiments was interpreted as an ohmic contribution of a quasi-normal part enriched with nonequilibrium quasiparticles. It have been shown that: 1) the polarity of the excess voltage depends on the polarity of the injection current; 2) the signal magnitude decreases with the thickness of a superconducting film and 3) the magnitude of the excess voltage tends to infinity when approaching the critical temperature. It is worth saying that due to technical constraints thin-film structures of the sandwich type studied in the seventies had a nonequilibrium superconductor (tin) thickness $w$ significantly smaller than the typical relaxation length of a charge imbalance $\Lambda_Q$, $w \ll \Lambda_Q$. Consequently, it was postulated that the nonequilibrium carrier concentration is constant along the whole thickness of the superconductor and a spatial relaxation dependence could be derived indirectly from a weak dependence on the thickness $\sim \exp(-w/\Lambda_Q) \simeq 1$. The majority of experiments were conducted at temperatures close to the critical one and the agreement with the phenomenological model \cite{Tinkham1972} has been established exactly in this high-temperature limit.

In numerous following works \cite{Schmid1975} qualitative conclusions of pioneer research \cite{Clarke1972} were confirmed using various materials and a simple phenomenological model \cite{Tinkham1972} was developed on more serious grounds \cite{Kopnin2001}. In particular it was shown that the current-voltage dependence of a normal metal-insulator-superconductor (NIS) detector $I(V)$ can be written as:
\begin{eqnarray}\label{CVC1}
I = \frac{G_{NN}}{e}\int\limits_{0}^{\infty} & \{N_S(E) [f_F(E) - f_F(E - eV) ] \nonumber
\\
& + [f_{k<} - f_{k>} ] \}dE,
\end{eqnarray}
where $G_{NN}$ is a tunneling conductance of a junction (detector) in a normal state, $f_F(E)$ is the Fermi distribution function, $N_S(E)$ is the density of states in superconductor, and $f_{k>}$, $f_{k<}$ are the populations of electron-like and hole-like branches of an excitation spectrum, correspondingly.

The magnitude of a charge imbalance $Q^{*}$ is defined exactly by the population difference of electron-like and hole-like branches,
\begin{equation}\label{Q}
Q^{*} = 2\int\limits_{0}^{\infty}N_{S}(E)[f_{k<} - f_{k>}]dE = -2N(0)\delta\mu_{S},
\end{equation}
where $N(0)$ is the single electron density of states at the Fermi level and $\delta\mu_S$ is the deviation of the chemical potential of Cooper pairs from its equilibrium value. It is worth mentioning that according to longitudinal and transverse excitation modes the expression in \Eq{CVC1} can be rewritten using nonequilibrium distribution functions $f_{T}^{N}$ and $f_{T}^{S}$, where symbols N and S are related to a normal metal and a superconductor electrode of a tunnel NIS detector,
\begin{equation}\label{CVC2}
I_d = \frac{G_{NN}}{e}\int\limits_{0}^{\infty}N_{S}(E)[f_{T}^{N} - f_{T}^{S}]dE,
\end{equation}
where
\begin{eqnarray}
&f_T(E, V_d, T_e) \equiv f_T^N(E, V_d, T_e) \nonumber
\\
&= \frac{1}{2} \tanh[(E + eV_d)/2k_BT_e] - \frac{1}{2} \tanh[(E - eV_d)/2k_BT_e]
\end{eqnarray}
is the equilibrium value of $f_{T}$ in the normal detector, $V_d$ is the detector voltage, $T_e$ is the electron temperature in the detector, and $f_{T}^{S}$ is the local nonequilibrium value of $f_{T}$ in the superconductor.

A simple comparison with the standard (equilibrium) equation for a tunnel current of a NIS junction at a temperature T \cite{Werthamer1966},
\begin{equation}\label{CVC3}
I(V, T) = \frac{G_{NN}}{e}\int\limits_{0}^{\infty}\left \{N_{S}(E)[f_F(E, T) - f_F(E - eV, T)] \right \}dE,
\end{equation}
allows us to conclude that all deviation from the equilibrium is defined by the last term in the integrand of \Eqs{CVC1} and \eref{CVC2}. The first term in the integrands of \Eqs{CVC1} and \eref{CVC2} corresponds to a standard (equilibrium) tunnel current \Eq{CVC3} which depends on the properties of the superconductor via the density of states. The energy dependence of the population differences of electron-like and hole-like branches of an excitation spectrum $[f_{k<} - f_{k>}]$ \textit{a priori} is unknown and, therefore, calculation of the dependence $I(V)$ for arbitrary displacement values of $V$ is not at all evident. However, the first term in the \Eq{CVC1} falls out at zero voltage and the so called excess current $I_{ex}$ is defined as with the help of \Eq{Q},
\begin{equation}\label{Iex}
I_{ex} \equiv I(V = 0) = \frac{G_{NN}}{e}\int\limits_{0}^{\infty}[f_{k<} - f_{k>}]dE = -\frac{G_{NN}}{e}\delta\mu_S.
\end{equation}
The equation \Eq{Iex} allows us to define the particularly microscopic parameter - the shift of a chemical potential of Cooper pairs $\delta\mu_S$ using easily measured in the experiment values $I_{ex}$ and $G_{NN}$.

There have recently appeared series of experimental papers where an attempt to develop the aforementioned observations for a case when an injection and a detection of nonequilibrium quasiparticles are spatially separated was made \cite{Yagi2003, Ikebuchi2004, Yagi2006, Yagi2009} (see Fig.~\ref{Yagi}). It has been indeed found that current-voltage characteristics of the tunnel NIS detector depend on the distance to injector NIS junction. Despite interesting observations, experiments \cite{Yagi2003, Ikebuchi2004, Yagi2006, Yagi2009} leave some open questions. For example, the relationship of a charge and energy imbalances was not investigated.

%%%%%%%%%%%%%%%%%%%%%%%%%%%%%%%%%%%%%%%%%%%%%%%%%%%%%%%%%%%%%%%%%%%%%%%%%%%%
%%%%%%%%%%%%%%%%%%%%%%%%%%%%%%%%%%%%%%%%%%%%%%%%%%%%%%%%%%%%%%%%%%%%%%%%%%%%

\section{Spatially resolved measurement setup}\label{setup}

\begin{figure}[tb]\label{Arutyunov}
\begin{center}
\epsfxsize=7cm\epsffile{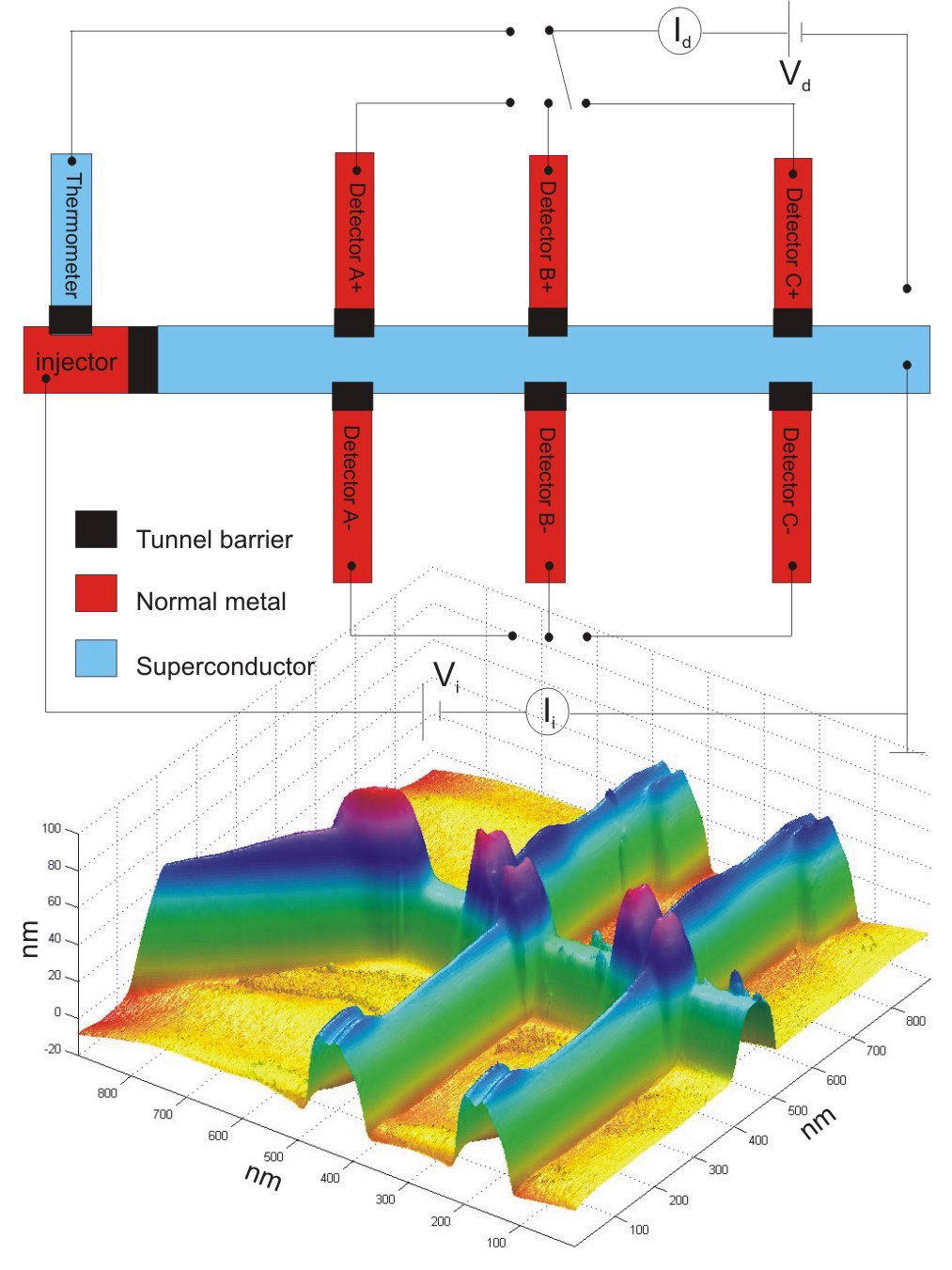}
\end{center}
\caption{Top panel: the design of a sample and measurements. The electron temperature of the injector made from a normal metal is measured by the NIS junction `thermometer'. The relaxation of nonequilibrium quasiparticles is measured either by a single NIS detector or by a pair of NIS junctions equidistant from the injector at the opposite sides of the superconductor (the NISIN configuration). Low panel: a scanning probe microscope image of the sample fragment. One can see the injector (the left massive copper electrode) and a pair of NIS detectors on the Si/SiO$_{x}$ substrate \cite{Arutyunov}.}
\end{figure}

In this section we discuss our experimental research of spatial dependencies of relaxations of the charge and energy imbalances in a superconductor at temperatures significantly lower than the critical temperature, $T \ll T_c$ \cite{Arutyunov}.

Multiterminal nanostructures were fabricated using electron beam lithography and ultra high vacuum evaporation of aluminum (superconductor, S) and copper (normal metal, N) separated by naturally-grown aluminum oxide (I) \cite{Arutyunov2001} (Fig.~\ref{Arutyunov}). The sample layout was
similar to the layout, described in \cite{Yagi2003, Ikebuchi2004, Yagi2006, Yagi2009} and
\cite{Yu1972, Dolan1977}. Electrons were injected from a normal metal through a tunnel junction into a superconductor. The typical thickness of the aluminum, detectors and the injector was 25, 40 and 80 $nm$, and the line width was 400, 180 and 1600 $nm$ respectively. The critical temperature of the aluminum microstrip was $T_c \simeq 1.35$ $K$ and electron mean free path was $l \simeq 20$ $nm$. The tunnel resistance of a large-area NIS injector was selected to be sufficiently low $\simeq 3$ $k\Omega $ in order to provide a high `pumping' quasiparticle current while a tunnel resistance of narrow NIS detectors was about $ 50$ $k\Omega $. In both cases the tunnel conductance of our junctions was low enough in order to neglect the proximity effect.

Experiments were performed in a $^3$He$^4$He dilution refrigerator located inside the electromagnetically shielded room using analog preamplifiers connected with the external measuring instruments through a system of radio-frequency filters. A typical measuring network consists of two circuits: an injector and a detector (Fig.~\ref{Arutyunov}, top panel). Two complementary configurations were used for the quasiparticle detection: a single NIS junction, or a pair of tunnel junctions (NISIN) which were equidistant from the injector. The second configuration was found to be more stabile with respect to parasitic potentials. However, it was obvious that the NISIN configuration cannot be used for the charge imbalance detection, since the signals from two NIS junctions compensate and extinguish each other due to the junctions reverse polarity with respect to the superconductor.

A special consideration was given to the thermalization of the system. In order to prevent the normal injector from overheating it was made sufficiently massive. An additional NIS electrode (marked as a `thermometer' in Fig.~\ref{Arutyunov} and placed at a distance of $\simeq$1 $\mu$m from the NIS injector) was used for measuring the electron temperature of the injector $T^i_e(I_i)$. The current-voltage characteristic of this `thermometer' $I_T(V_T, I_i = const)$ was measured for fixed injector current values $I_i$ and the electron temperature was calculated from the fitting of experimental current-voltage characteristics by theoretical curves for the NIS junction when the electrode disequilibrium could be neglected, \Eq{CVC3}. After calibrating the injector the dependence of the electron temperature on the injection current has been established $T_e^i (I_i, T = const)$, and the thermometer circuit was disconnected for the rest of experiments. Here and below the $I_T(V_T)$, $I_i(V_i)$ and $I_d(V_d)$ dependencies denote current-voltage characteristics of the thermometer, the injector and the detector, respectively.

%%%%%%%%%%%%%%%%%%%%%%%%%%%%%%%%%%%%%%%%%%%%%%%%%%%%%%%%%%%%%%%%%%%%%%%%%%%%
%%%%%%%%%%%%%%%%%%%%%%%%%%%%%%%%%%%%%%%%%%%%%%%%%%%%%%%%%%%%%%%%%%%%%%%%%%%%

\section{Temperature parameters}\label{temperatures}

\begin{figure}[tb]\label{Arutyunov2}
\begin{center}
\epsfxsize=7cm\epsffile{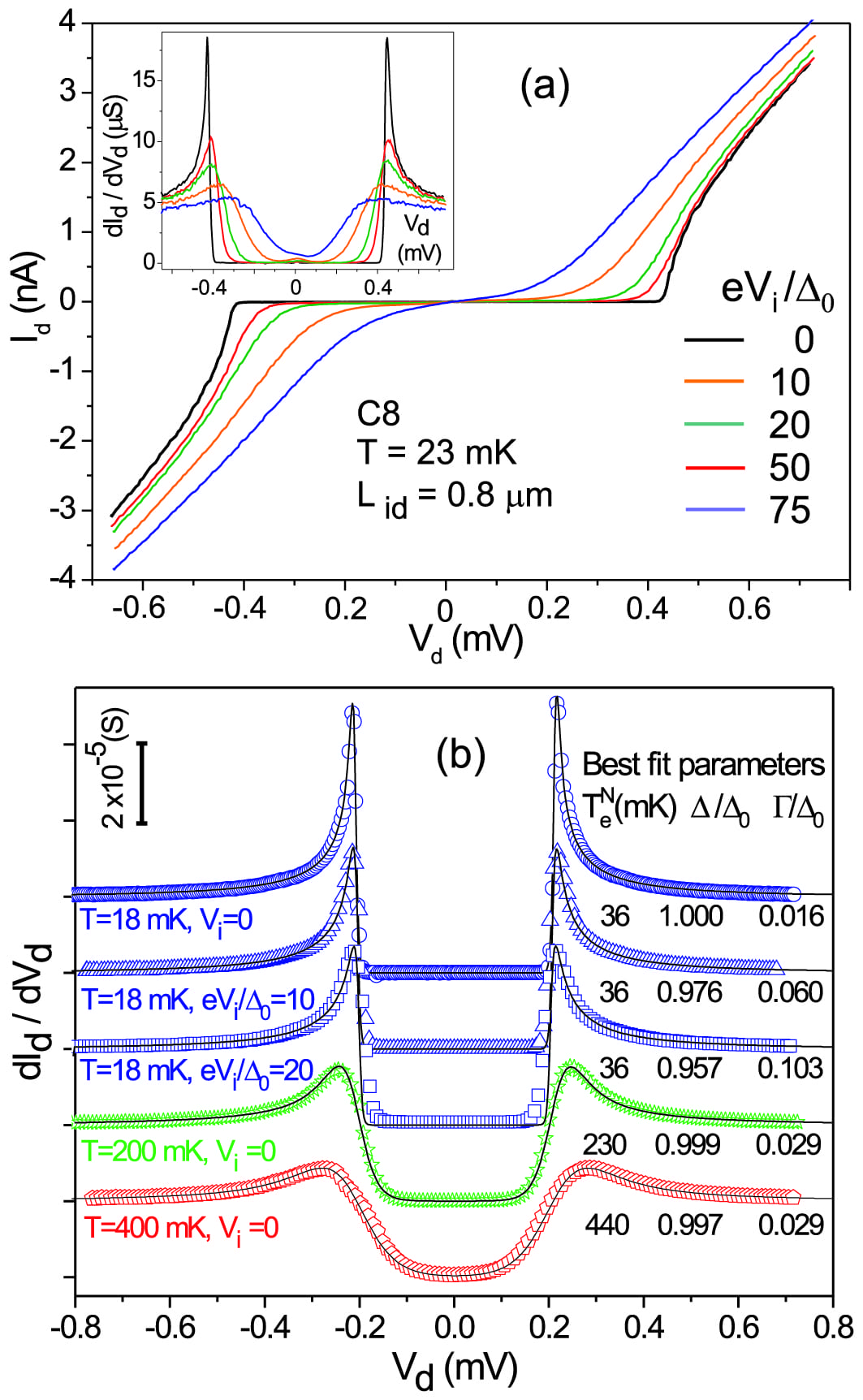}
\end{center}
\caption{(a) Sample C8: experimental current-voltage characteristics of the double junction NISIN detector located at the distance $L_{id} = 0.8$ $\mu$m from the injector at various pumping energies $eV_i$. Inset: the corresponding $d I_d/d V_d$ dependencies obtained by a phase-sensitive detector measurements of the first harmonic of the signal, modulated by the 19 Hz frequency. A weak asymmetry which can be seen at a strong injection probably results from a nonideal symmetry of the NIS junctions connected in series. (b) Sample C2: first derivatives of the current-voltage characteristics of the NISIN detector located at the distance $L_{id} = 2.8$ $\mu$m from the injector and measured at the temperature $T = 18$ mK and at three pumping energies $eV_i/\Delta_0 = 0$ $(\bigcirc)$, $10$ $(\bigtriangleup)$ and $20$ $(\Box)$; and at $T = 200$ mK $(\bigstar)$ and $400 $ mK $(\diamondsuit)$ at zero injection. Solid lines correspond to the theoretical fits with the fitting parameters presented in the figure \cite{Arutyunov}.}
\end{figure}

It is instructive to summarize the definitions of various parameters of the dimensionality `temperature' hereafter used in this review.
The parameter $T$ corresponds to the bath temperature measured by the two resistors made of RuO$_x$, which are thermally and mechanically reliably connected to the walls of the mixing chamber and to the massive copper sample holder. Both sensors were calibrated by a nuclear orientation thermometer and during the measurements their readings deviated less than by a couple of mK.

$T_e^i$ and $T_e^d$ denote electron temperatures of the injector ($i$) and the detector ($d$) made from a normal metal (copper). Both values were defined from the fitting of the experimental $I_i(V_i)$ and $I_d(V_d)$ dependencies by an expression for the tunnel current of the NIS junction when nonequilibrium effects could be neglected, given by \Eq{CVC3}. The electron temperature of the injector depends on the injection current $T_e^i(I_i, T = const)$ due to the Joule heating. Despite sufficiently low currents used in the experiments this trivial phenomenon can lead to tangible overheating at ultralow temperatures. At the lowest bath temperatures $T \simeq 20 $ $mK$ and maximal injection currents $I_i \simeq 1$ $\mu A$
the overheating of the injector subsystem was $\delta T_e^i \equiv T_e^i - T \simeq$ 100 mK. It should be noted that in the discussed experiments $T_e^i$ is the highest `real' temperature of the whole sample. Several parameters with the `temperature' dimension will sufficiently exceed $T_e^i$ which is the evidence of their `effective' nature, as it will be shown below.

The electron-phonon interaction is very week at ultralow temperatures \cite{Giazotto2006}, and a heat conductivity of a long superconducting sample part is exceptionally low. Therefore, the heat spreading effect from the `hot' injector along the superconducting sample should be small enough.
Subsequently, it is reasonable to consider that the phonon temperature of the remote detectors should not significantly deviate from the bath temperature $T$. On the contrary, the electron temperature of the normal metal lead of an NIS detector $T_e^d$, determined by fitting the experimental $I_d(V_d, I_i = 0)$ dependencies at the zero injection current by the \Eq{CVC3}, is always lower than the bath temperature $T$.
This phenomenon is typical for galvano-magnetic measurements at ultra-low temperatures and is related to the inevitable heating of the electronic subsystem by the electromagnetic radiation which enters through junction leads. This heating channel can be reduced with the help of different kinds of the high frequency filtration but can never be completely eliminated. At the lowest bath temperatures $T \simeq$ 20 mK and zero injection currents $I_i \simeq 0$, an increase of electron temperature $\delta T_e^d \equiv T_e^d - T$ in NIS detectors varied from 10 to 40 mK for different junctions.
This is a quite worthy result which is the evidence of the quality of the used high-frequency filters. Since each detector is isolated from the `hot' injector by two tunnel NIS junctions divided by the extended superconducting sample `body' which has a low heat conductivity and a good thermal connection with the substrate (Si/SiO$_x$), it is  reasonable to assume that the electron temperature of the detector does not depend on the injection current $I_i$ and is defined only by the level of the electromagnetic noise, which reaches a specific electrode, $T_d^e(I_i > 0, T) \simeq T_d^e(I_i = 0, T)$.

The $T_e^S$ denote the superconductor electron temperature. Strictly speaking, $T_e^S$ should be determined from a complicated energy balance equations \cite{Giazotto2006}, which parameters, for example a metal-substrate boundary heat conductivity, are usually not accurately defined.
From the very general considerations it can be expected that the thermodynamic electron temperature in a superconductor, which enters into the nonequilibrium distribution function $f_{T}^S (E, T_e^S)$, should be higher than the phonon temperature $T_{phonon} \approx T$ at finite injection currents $I_i > 0$. However, it should be immediately noted that the $T_e^S$ parameter cannot be determined in the context of the phenomenological formalism employed in this review. Deeper microscopic approach (which is as far as we know currently absent) should be used to determine the nonequilibrium and essentially asymmetric with respect to the chemical potential distribution function $f^S_{T} (E, T_e^S)$. Otherwise, any arbitrary value $T_e^S$ substituted in a symmetric function $f_{T}^{S} (E, T_e^S)$, for example the equilibrium function,  gives exactly the same tunnel current of an NIS junction.

Finally, the last parameter with the `temperature' dimension, $T^{\ast}$, characterizes the superconductor energy gap $\Delta$. As it will be shown further, the injection of nonequilibrium quasiparticles into a superconductor among other effects leads to a gap supression. Therefore two alternative descriptions are possible: either to obtain the dependence $\Delta(I_i)$ directly from the experimental data, or the gap suppression can be `converted' into the effective temperature $T^{\ast}$ using the standard (equilibrium) BSC theory. In the latter case the $T^{\ast}$ parameter indicates which equilibrium temperature $T = T^{\ast}$ corresponds to the nonequilibrium gap $\Delta(I_i > 0) = \Delta_{BSC}(I_i = 0, T^{*})$.
Obviously, the $T^{\ast}$ value is only a convenient parameter for the description of a gap supression and has no direct link with the `true' thermodynamic temperature $T_e^S$. As it will be shown further the injection of nonequilibrium charge carriers into a superconductor can lead to various phenomena. In particular, the longitudinal mode (the energy imbalance) cannot be fully described only by the gap suppression or, alternatively, by the rise of the effective temperature $T^{\ast}$.

%%%%%%%%%%%%%%%%%%%%%%%%%%%%%%%%%%%%%%%%%%%%%%%%%%%%%%%%%%%%%%%%%%%%%%%%%%%%%%%%%%%%%%%%%%%%%%%
%%%%%%%%%%%%%%%%%%%%%%%%%%%%%%%%%%%%%%%%%%%%%%%%%%%%%%%%%%%%%%%%%%%%%%%%%%%%%%%%%%%%%%%%%%%%%%%

\section{Nonequilibrium quasiparticle injection}\label{injection}

%%%%%%%%%%%%%%%%%%%%%%%%%%%%%%%%%%%%%%%%%%%%%%%%%%%%%%%%%%%%%%%%%%%%%%%%%%%%%%%%%%%%%%%%%%%%%%%
%%%%%%%%%%%%%%%%%%%%%%%%%%%%%%%%%%%%%%%%%%%%%%%%%%%%%%%%%%%%%%%%%%%%%%%%%%%%%%%%%%%%%%%%%%%%%%%

%
\begin{figure}[t]
\begin{center}
\epsfxsize=7cm\epsffile{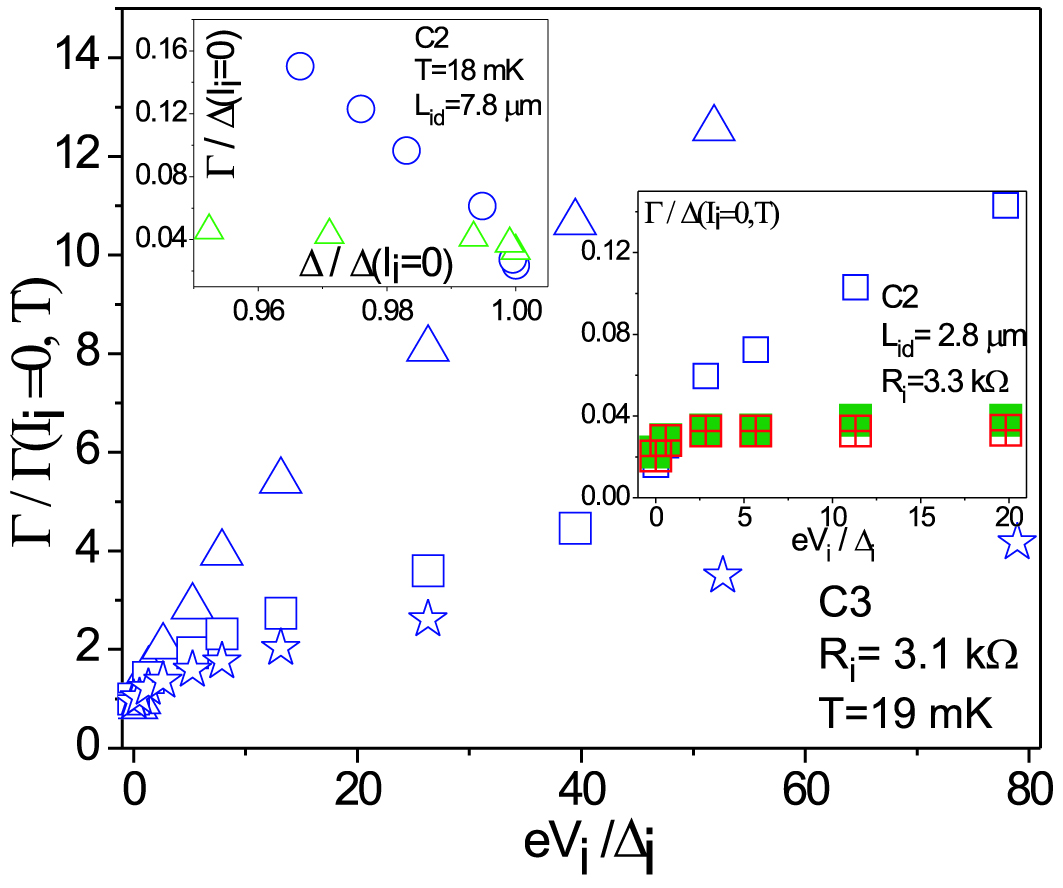}
\end{center}
\caption{Sample C3: the dependence of the DOS smearing parameter $\Gamma$ on the injection energy $eV_i$ for three detectors located at the distances $L_{id} = 0.8$ $(\bigtriangleup)$, $2.8$ $(\Box)$ and $ 7.8$ $(open star)$ $\mu$m from the injector. Left inset: the dependence of the DOS smearing parameter $\Gamma(I_i)$ on the energy gap $\Delta(I_i)$ measured at the equal injection currents $I_i$ on the sample C2 using the NIS detector located at at the distance $L_{id} = 7.8$ $\mu$m at temperatures $T = 18$ mK $(\bigcirc)$ and $T = 200$ mK $(\bigtriangleup)$. Rright inset: the sample C2, the dependence of the DOS smearing parameter $\Gamma$ on the injection energy obtained on the same detector $L_{id} = 2.8$ $\mu$m at different temperatures $T = 17.5$ $(\Box)$, $200$ $(\blacksquare)$ and $400$ $(\boxplus)$ mK \cite{Arutyunov}.}%
\label{Arutyunov3}%
\end{figure}

The nonequilibrium quasi-particle injection into a superconductor can lead to a deviation of the density of states (DOS) $N_S(E)$ and the distribution functions $f_{L, T}^S(E, T_e^S)$ from their equilibrium values as it was discussed earlier. The distribution function can be found by the deconvolution of the experimental tunnel current-voltage characteristics under certain conditions \cite{Pothier1997, Pierre2001}.

In general case, in order to find the distribution function of a superconductor the self-consistent solution of a Keldysh-Usadel equation is needed \cite{Belzig1999}, which is an exceptionally complicated task. We are going to use a simplified approach for an interpretation of our results by postulating the identity of functional forms of the distribution function and the density of states with their equilibrium values,
\begin{eqnarray}
f_S(E, t_e^S, \mu_s) = \frac{1}{\exp[(E - \mu_S)/k_B T_e^S] + 1},\label{f_S}
\\
N_S(E, \Delta, \Gamma) = \Big\vert \frac{\re(E + i\Gamma/2)}{\sqrt{(E + i\Gamma/2)^2 - \Delta^2}}\Big\vert.\label{N_S}
\end{eqnarray}

The deviation of $f_L^S$ from its equilibrium value formally corresponds to an increase of the temperature $T_e^S$ above the bath temperature $T$, while the deviation of $f_T^S$ corresponds to a finite value of the chemical potential $\mu_S$ measured with the respect to the Fermi level.
Formally, if the expression \Eq{f_S} correspond nominally to the reality, the three parameters $T_e^S$, $\mu_S$ and $\Gamma$ would be sufficient for describing the nonequilibrium state of the superconductor. However, as mentioned previously, no symmetric function, including the Fermi function \Eq{f_S}, cannot be even qualitatively used for determining the tunnel current of the NIS junction where a superconducting electrode is in a nonequilibrium state. The only role using the inapplicable expression \Eq{f_S} is to obtain a finite value of a chemical potential $\mu_S$, which is physically connected with the charge imbalance. For the analysis of the experimental data we are going to use three following fitting parameters in our phenomenological approach. The $\Gamma$ parameter, which defines the broadening of the density of states \cite{Dynes1978, Dynes1984}, the superconductor energy gap $\Delta$, and the effective chemical potential $\mu_S$ \cite{Owen1972, Parker1975}.
All these parameters can be defined from the experimental $I_d(V_d)$ and $d I_d/d V_d$ dependencies. It is worth mentioning once again that the thermodynamic temperature of a superconductor $T_e^S$ cannot be defined in the framework of this phenomenological approach.

Since the injected nonequilibrium quasiparticles have to relax at certain times (or, equivalently, at certain distances), it should be expected that the fitting parameters have to depend both on a pumping level (of the energy $eV_i$ or the current $I_i = V_i G_{NN}^i$) and on the distance between the detector and the injector $L_{id}$. The $I_{d}(V_{d})$ dependence of a NIS detector in the presence of nonequilibrium injection is given by \Eq{CVC2}.

The shape of the current-voltage characteristic $I_d(V_d)$ depends on the quasiparticle injection level (Fig.~\ref{Arutyunov2}). By fitting the experimental $I_d(V_d)$ and $dI_d/dV_d$ data the relevant energy, temperature and spatial dependencies $\Gamma(eV_i, T, L_{id})$ and $\Delta(eV_i, T, L_{id})$ can be determined (Fig.~\ref{Arutyunov3}, \ref{Arutyunov4}). The smearing of the current-voltage characteristic at values of bias close to the gap [the smothering of `corners' at $eV_i \simeq \Delta (I_i)$] is determined by the $\Gamma (I_i)$  parameter included in the expression for the superconductor density of states \Eq{N_S} and related to the finite lifetimes of quasiparticle excitations \cite{Dynes1978, Dynes1984}.
For a given injection current $I_i$ the temperature dependence $\Gamma(T, I_i = const, L_{id} = const)$ (right inset in Fig.~\ref{Arutyunov3}) most probably originates from the more intensive quasiparticle relaxation due to the inelastic scattering on phonons.
The total number of nonequilibrium quasiparticles injected into the superconductor lead per unit time is proportional to the injection current $I_i$ (or alternatively to the injection energy $eV_i$) but only part of them $\sim\exp(-L_{id}/\sqrt{\mathcal{D}\tau_{r}})$ reaches the detector located at a distance of $L_{id}$ from the injector. Here $\mathcal{D} = \frac{1}{3} v_F \ell$ is the diffusion coefficient, and
$v_F$ is the Fermi velocity. At low temperatures the quasiparticle recombination time $\tau_r$ can be estimated as
\begin{equation}
\tau_{0}/\tau_{r} \sim (T/T_{c})^{1/2}\exp(-\Delta/k_{B}T),
\end{equation}
where $\tau_0$ is a characteristic time of the electron-phonon scattering \cite{Kaplan1976, Chi1979}.
These rather simple considerations qualitatively explain the observed spatial, energy and temperature dependencies of the splitting parameter $\Gamma$ (Fig.~\ref{Arutyunov3}). However, microscopic models are required for the quantitative analysis which have to include the contribution of the electromagnetic environment \cite{Pekola2010}.

%%%%%%%%%%%%%%%%%%%%%%%%%%%%%%%%%%%%%%%%%%%%%%%%%%%%%%%%%%%%%%%%%%%%%%%%%%%%%%%%%%%%%%%%%%%%%%%
%%%%%%%%%%%%%%%%%%%%%%%%%%%%%%%%%%%%%%%%%%%%%%%%%%%%%%%%%%%%%%%%%%%%%%%%%%%%%%%%%%%%%%%%%%%%%%%

\section{Energy imbalance (longitudinal mode)}\label{energy}

%%%%%%%%%%%%%%%%%%%%%%%%%%%%%%%%%%%%%%%%%%%%%%%%%%%%%%%%%%%%%%%%%%%%%%%%%%%%%%%%%%%%%%%%%%%%%%%
%%%%%%%%%%%%%%%%%%%%%%%%%%%%%%%%%%%%%%%%%%%%%%%%%%%%%%%%%%%%%%%%%%%%%%%%%%%%%%%%%%%%%%%%%%%%%%%

\begin{figure}[t]
\begin{center}
\epsfxsize=7cm\epsffile{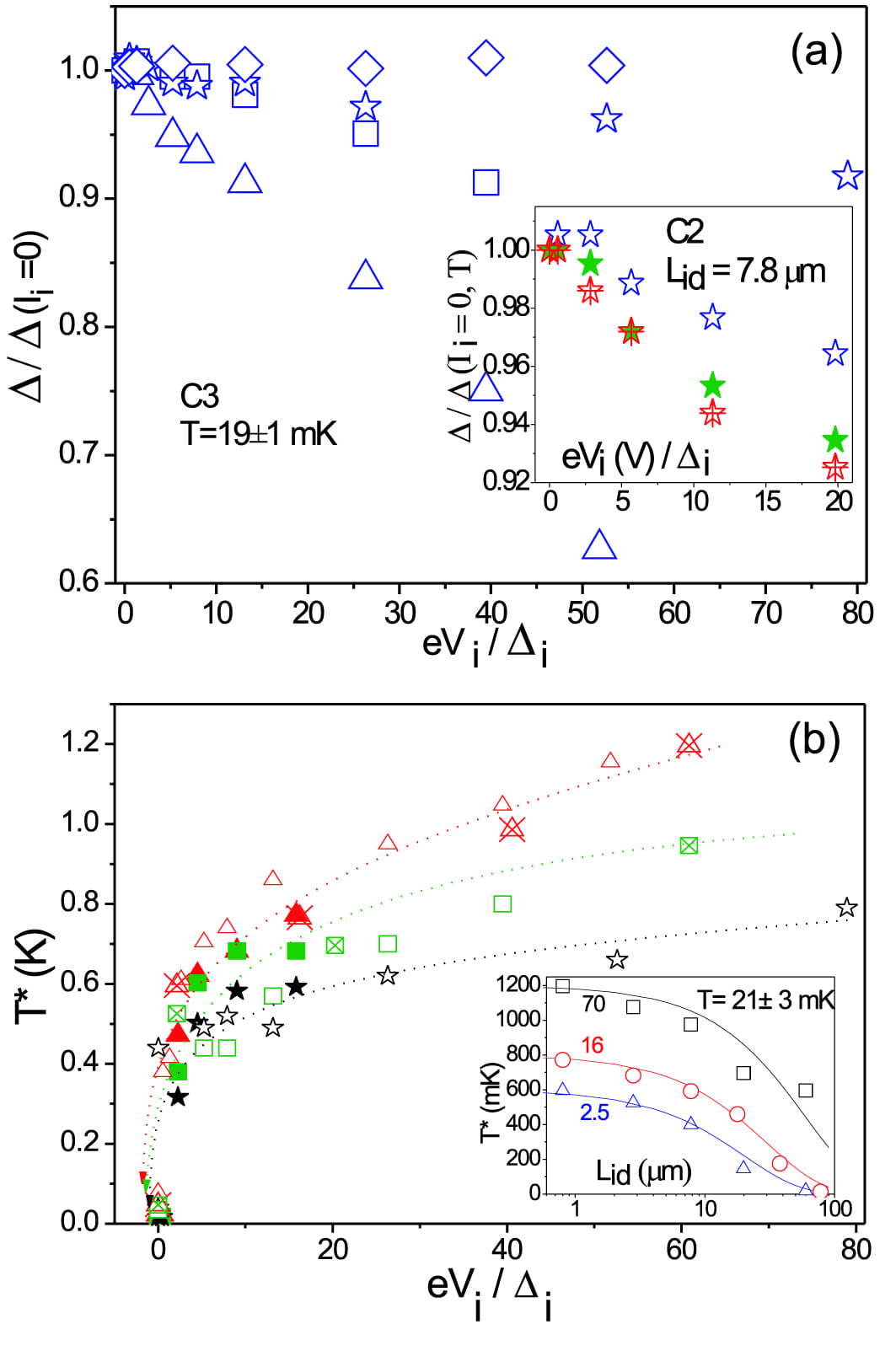}
\end{center}
\caption{(a) Sample C3: the dependencies of the energy gap $\Delta$ on the pumping energy $eV_i$ measured at different distances from the injector  $L_{id} = 0.8$ $(\bigtriangleup), 2.8$ $(\Box)$, $ 7.8$ $(open star)$ and $17.8$ $(\diamondsuit)$ $\mu$m. Inset: the same dependencies measured by the NIS detector at the distance $L_{id} = 7.8$ $\mu$m at different temperatures: $T = 17.5$ (open star), $200$ $(\bigstar)$ and $400$ (crossed star) mK. (b) Dependencies of the effective temperature $T^{\ast}$ on the injection energy $eV_i$ measured on different samples: C3 (18 mK, hollow symbols), C2 (19 mK, filled symbols) and C8 (24 mK, crossed symbols) located at the distances $L_{id} = 0.8$ $(\bigtriangleup)$, $2.8$ $(\Box)$ and $7.8$ $(\bigstar)$ $\mu$m. Inset: spatial dependencies of the effective temperature $T^{\ast}$ averaged over several samples at injection energies $eV_i/\Delta_i = 70$ $(\Box)$, $16$ $(\bigcirc)$ and $2.5$ $(\bigtriangleup)$. Lines are fits with the exponential dependence $\sim \exp(-L_{id}/\Lambda_{T^{*}})$ \cite{Arutyunov}.}%
\label{Arutyunov4}
\end{figure}

Neglecting the phase of the superconductor order parameter, the gap equation for the pairing potential $\Delta$ to be solved self-consistently with the Keldysh-Usadel equation, is
\begin{equation}
\Delta=\lambda\int\limits_{0}^{\infty}f_{L}^{S}(E,L_{id})\re(F)dE,
\end{equation}
where $\lambda$ is the electron-phonon interaction constant, $F$ is the anomalous Green's function (pair amplitude), and $f_L^S (E,L_{id})$ is the local  nonequilibrium value of the longitudinal component $f_{L}$ of the nonequilibrium distribution function \cite{Belzig1999}.
For any arbitrary energy value $E$ the function $f_{L}^{S}(E,L_{id})$ is always smaller than its equilibrium value $\tanh(E/k_BT)$ \cite{Vasenko2009}, thereby reducing $\Delta$ [Fig.~\ref{Arutyunov4}(a)]. This effect increases with the bath temperature rise [Fig.~\ref{Arutyunov4}(a), inset].
Using the well-known temperature dependence of the BCS model we can describe the gap supression due to the quasiparticle injection as the rise of some effective temperature $T^{\ast}$, $\Delta(T_{e}^{S},I_{i},L_{id})=\Delta_\mathrm{BCS}(T^{\ast},I_{i}=0,L_{id})$ [Fig.~\ref{Arutyunov4}(b)].
It is reasonable to assume that together with the spatial relaxation of the injected quasiparticles the gap supression (or alternatively the rise of the $T^{\ast}$) should also decay at a certain characteristic distance $\Lambda_{T^{\ast}}$,
\begin{equation}
T^{\ast} = T^{\ast}(0)\exp(-L_{id}/\Lambda_{T^{\ast}}).
\end{equation}

The experimental data analysis results in a huge (on the microscopic scale) value of the characteristic length $\Lambda_{T^{\ast}}=$ 40 $\mu$m $\pm$ 20 $\mu$m [Fig.~\ref{Arutyunov4}(b), inset]. A relatively big error $\sim 50 \%$ is related to the weak temperature dependence $\Delta_{BCS}(T)$ at low temperatures $T \ll T_c$ and to a limited number of experimental points given by the quantity of NIS detectors in tested samples (Fig.~\ref{Arutyunov}). It is worth mentioning once again that $T^{\ast}$ is nothing more than a convenient parameter which has no direct relation to the thermodynamic temperature of the superconductor, $T_e^S$. It should be noted that in the limit of strong quasiparticle pumping and weak electron-electron interaction a highly nonequilibrium state can be realized. In this case the thermodynamic temperature $T_e^S$ introduction is quite problematic. However, the energy gap value $\Delta (I_i)$ or, alternatively $T^{*} (I_i)$ can be obtained from experimental current-voltage characteristics $I_d(V_d, I_i = const)$.

We assume the energy gap supression $\Delta(I_i)$ (Fig.~\ref{Arutyunov4}) and the density of states smearing $\Gamma(I_i)$ (Fig.~\ref{Arutyunov3}) to be the manifestations of the same phenomenon - the energy imbalance. It is likely that in a more general (microscopic) model both excitation modes of nonequilibrium charge carries (the longitudinal and the transverse) are going to be entangled and consistently describe the dependencies $\Delta(I_i)$ and $\Gamma(I_i)$. Within the framework of the standard BCS theory the current-voltage characteristic of the NIS detector junction [expression \Eq{CVC2}] is defined exclusively by the distribution function $f_N(E, V, T_{e}^{N})$ in the normal electrode and depends on the parameters of the superconductor only via the value of the energy gap $\Delta$. The temperature of the superconductor $T_e^S$ contributes to the current-voltage characteristic only through the temperature dependence $\Delta(T_e^S)$. In an equilibrium state $T_e^S = T_e^N = T_{phonon} = T$ the experimentally observed smearing of the current-voltage characteristic of the NIS junction is only due to the finite temperature of the normal electrode. Our approach for the interpretation of the experimental data differs from the standard BCS theory due to the introduction of the splitting parameter $\Gamma(I_i)$ included in the expression for the density of states \Eq{N_S}, and contributed to the additional (not thermal) smearing of the current-voltage $I_{d}(V_{d})$ dependence, \Eq{CVC2}. The use of the `equilibrium' tunnel current expression [first term in \Eq{CVC2}] with the modified density of states [\Eqs{f_S} and \eref{N_S}] allows us to separate the influence of two contributions, $\Delta(I_i)$ and $\Gamma (I_i)$.
The gap suppression $\Delta (I_i)$ by the finite injection current (Fig.~\ref{Arutyunov4}) leads to a `sharp' current-voltage characteristics of the detectors $I_d(V_d, I_d > 0)$, but with a smaller value of the gap $\Delta (I_i > 0) < \Delta (I_i = 0)$. The contribution of the finite smearing of the density of states $\Gamma (I_i)$ manifests itself as the `smoothing' of the current-voltage characteristics at voltages $eV_d \simeq \pm \Delta(I_i)$, which cannot be explained by increase of normal electrode temperature, $T_e^N$, within reasonable limits.

The current-voltage characteristics of the detectors measured at zero injection currents can be sufficiently accurately described by the equilibrium expression for the NIS junction current \Eq{CVC3} under the assumption of the finite but rather small Dynes smearing of the density of states $\Gamma(I_i = 0)/\Delta(I_i = 0) \simeq 0.02$ [Fig.~\ref{Arutyunov2}(b)] in agreement with the existing literature data \cite{Pekola2010}. However, at finite injection currents the experimental current-voltage characteristics $I_d(V_d, I_i > 0, T)$ cannot be described within the standard BCS theory taking into account that corresponding values have increased above the bath temperature, $T_e^S, T_e^N \gg T$. At sufficiently low temperatures $T' \ll T_c$ the shape of the experimental current-voltage characteristics $I_d(V_d, I_i > 0, T')$ is qualitatively different from the equilibrium dependencies obtained at higher temperatures $I_d(V_d, I_i = 0, T'' > T')$ [Fig.~\ref{Arutyunov2}(b)]. If we assume at a bath temperature $T$ that the current-voltage characteristic broadening is explained solely by the rise of the effective temperature $T_e^S = T^{*} > T$  than the same temperature value should be also substituted into the BCS dependence $\Delta (I_i, T) = \Delta_{BCS}(I_i = 0, T_e^S)$. However, it turns out that following this description, it should be assumed that the temperature of the superconductor is considerably higher than the `hottest' system point - the normal injector: $T_e^S = T^{\ast} (I_i) \gg T_e^{i}(I_i)$ which contradicts the common sense.

By summarizing these arguments we come to the conclusion that for the description of the experimental current-voltage characteristics at finite injection currents within the phenomenological model the introduction of the effective temperature $T_e^S$ is not sufficient, and consequently, it is necessary to use two independent parameters, $\Delta(I_i)$ and $\Gamma(I_i)$. The dependencies of these parameters on the injection current are different. The effect of the gap suppression is sufficiently small (just a few percentages) at significantly low temperatures $T \ll T_c$ and not so big injection energies $eV_i/\Delta_0 \lesssim 10$ [Fig.~\ref{Arutyunov4}(a)]. At the same time the rise of the parameter $\Gamma$ is quite significant at the same injection levels (Fig.~\ref{Arutyunov3}). With the temperature increase the $\Gamma(I_i)$ dependence became weaker (insets in Fig.~\ref{Arutyunov3}). It is worth to emphasize that the introduction of two parameters $\Delta(I_i)$ and $\Gamma(I_i)$ is the result of our phenomenological approach, which postulates that the current-voltage characteristics of the detectors can be described by the standard expression for the NIS junction tunnel current \Eq{CVC3} with the superconductor density of states given by \Eqs{f_S}, \eref{N_S}. We hope that with further development of the nonequilibrium superconductivity theory the whole variety of the experimental data could be described by using just a single parameter, the thermodynamic temperature $T_e^S$, included into a nonequilibrium distribution function $f_S (E, T_e^S)$.

%%%%%%%%%%%%%%%%%%%%%%%%%%%%%%%%%%%%%%%%%%%%%%%%%%%%%%%%%%%%%%%%%%%%%%%%%%%%%%%%%%%%%%%%%%%%%%%
%%%%%%%%%%%%%%%%%%%%%%%%%%%%%%%%%%%%%%%%%%%%%%%%%%%%%%%%%%%%%%%%%%%%%%%%%%%%%%%%%%%%%%%%%%%%%%%

\section{Charge imbalance (transverse mode)}\label{charge}

%%%%%%%%%%%%%%%%%%%%%%%%%%%%%%%%%%%%%%%%%%%%%%%%%%%%%%%%%%%%%%%%%%%%%%%%%%%%%%%%%%%%%%%%%%%%%%%
%%%%%%%%%%%%%%%%%%%%%%%%%%%%%%%%%%%%%%%%%%%%%%%%%%%%%%%%%%%%%%%%%%%%%%%%%%%%%%%%%%%%%%%%%%%%%%%

%
\begin{figure}[t!]\label{imbalance}
\begin{center}
\epsfxsize=5cm\epsffile{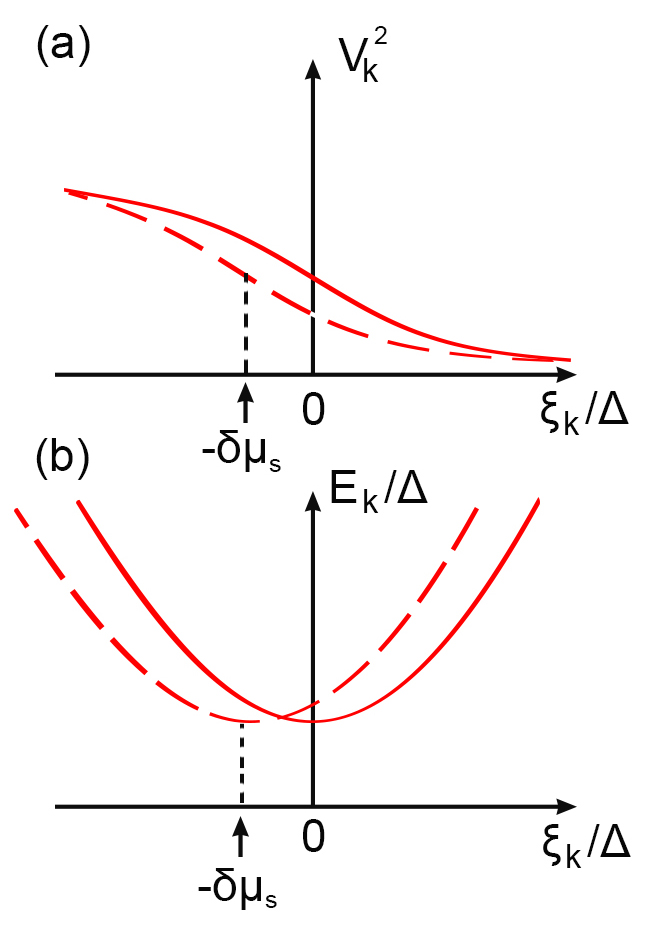}
\end{center}
\caption{The influence of the charge imbalance with the excess of electron-like quasiparticles on the condensate density (a) and the excitation spectrum (b). Solid lines correspond to the equilibrium state, dashed lines - the nonequilibrium state.}
\end{figure}

\begin{figure}[t!]\label{Arutyunov5}
\begin{center}
\epsfxsize=7cm\epsffile{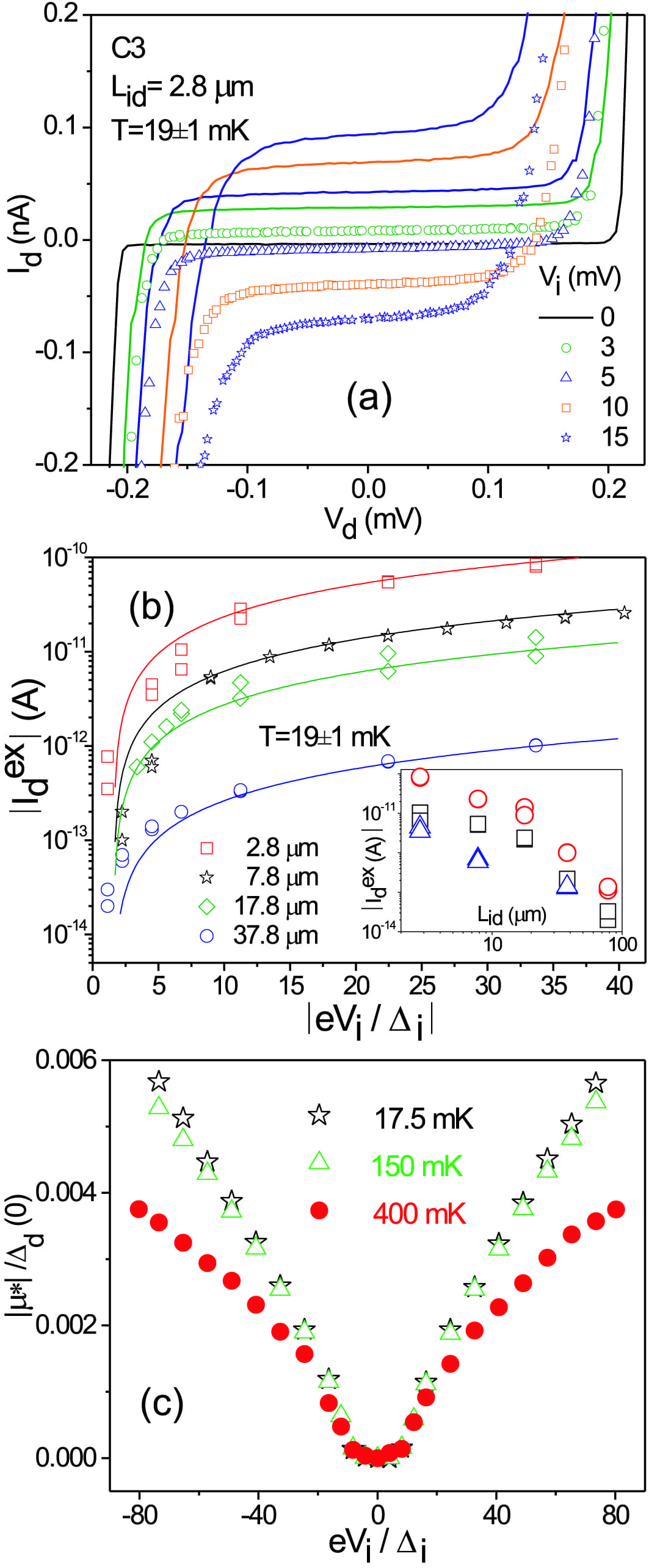}
\end{center}
\caption{Sample C3. (a) Fragments of the experimental current-voltage characteristics at bias shifts less than the gap $eV_d \simeq \Delta$ for the NIS detector located at the distance $L_{id} = 2.8 \mu$m from the injector at various injection levels shown in the figure. Lines correspond to the same injection values but with the different polarity. (b) The dependency of the detector excess current $I_d^{ex}$ on the injection energy $eV_i/\Delta_i$ for different distances between the detector and the injector. Lines are the calculations which use the parameter $\Lambda_{Q^{\ast}} = 3.6, 3.6, 5.2$ and $6.4$ $\mu$m for the detectors with $L_{id} = 2.8, 7.8, 17.8$ and $37.8$ $\mu$m respectively. Inset: the spatial dependence of the excess current measured at injection energies $\vert eV_i/\Delta_i \vert = 33.8$ $(\bigcirc)$, $6.7$ $(\Box)$ and $4.5$ $(\bigtriangleup)$. (c) The dependence of the chemical potential $\mu^{\ast} \equiv \mu_S$ on the injection energy measured at different temperatures by the same NIS detector at the distance $L_{id} = 7.8$ $\mu$m from the injector \cite{Arutyunov}.}
\end{figure}

Lets now move on to the analysis of another interesting phenomenon manifested at the same experiments on the injection of nonequilibrium quasiparticles - the charge imbalance. As was discussed above, see \Eq{CVC1}, the population difference between the electron-like and the hole-like spectrum branches of the excitation spectrum $f_{k>} \neq f_{k<}$ leads to the appearance of the finite NIS detector current at a zero $V_d$ referred to as the excess current, $I^{ex}_d \equiv I_d(V_d = 0)$. This effect has been observed both in early works on flat structures of the `sandwich' type \cite{Clarke1972, Tinkham1972} and in the recent research on multiterminal NIS structures \cite{Yagi2003, Ikebuchi2004, Yagi2006, Yagi2009}.
Within the phenomenological description the nonzero value of the excess current is related to the deviation of the Cooper pairs chemical potential $\mu_S$ from its equilibrium value [expression \Eq{Iex}], see Fig.~\ref{imbalance}. The typical example of the experimental current-voltage characteristic of the NIS detector at the voltage shifts smaller than the gap, $eV_d \simeq \Delta$, at various quasi-particle injection levels is shown in Fig.~\ref{Arutyunov5} (a). As it was expected, the excess current $I_d^{ex}$ increases with the pumping energy rise and changes the sign with the change in the polarity of the injection current. The effect weakens with increasing distance between the detector and the injector $L_{id}$ [Fig.~\ref{Arutyunov5} (b)] and with the bath temperature rise [Fig.~\ref{Arutyunov5} (c)]. The weak asymmetry of the excess current with respect to the zero injection current [Fig.~\ref{Arutyunov5} (a)] is probably related to the presence of parasitic residual thermo-electric potentials in the measuring circuit due to significant temperature gradients between the preamplifiers (at a room temperature) and the sample.

Once again we would like to note that in NISIN detector geometry (Fig.~\ref{Arutyunov}) the excess current $I_d^{ex}$ is absent due to the signal compensation from the two NIS junctions connected with the opposite polarity with respect to the superconductor. The NISIN geometry was
used only as a complementary configuration in the experiments on the energy mode relaxation [Fig.~\ref{Arutyunov2}(a)].

The solution of the diffusion equation for the charge imbalance $Q^{\ast}$ leads to the expression for the excess current:
\begin{equation}
I_{d}^{ex}=I_{i}\frac{F^{\ast}\Lambda_{Q^{\ast}}G_{NN}}{2e^{2}N(0)\mathcal{D}\sigma
}\exp(-L_{id}/\Lambda_{Q^{\ast}}),
\end{equation}
where $N(0)=1.08\times10^{47}$ 1/J$\times$m$^{3}$ is the density of state of the aluminum at the Fermi level (in a normal state), $F^{\ast}$ is a slowly varying function which is equal to zero at $V_i = 0$ and equal to unity at $eV_i \gg \Delta_i$ \cite{Tinkham1972}. The substitution of corresponding values: the mean free path $\ell\simeq20$ nm, the Fermi velocity $v_{F}=1.36\times10^{6}$ m/s, the diffusion coefficient $\mathcal{D}=\frac{1}{3}v_{F}\ell$, and the cross-section of the superconducting channel $\sigma=25$ nm $\times$ $400$ nm provides us with the satisfactory agreement with the experimental dependencies $I_d^{ex}(e V_i)$ [Fig.~\ref{Arutyunov5}(b)]. The obtained value for the relaxation length of the charge imbalance $\Lambda_{Q^{\ast}}=\sqrt{\mathcal{D}\tau_{Q^{\ast}}}$ varies from 3.5 to 6.5 $\mu$m being in a reasonable agreement with the existing data \cite{Yagi2003, Ikebuchi2004, Yagi2006, Yagi2009, Hubler2010}.

A weak temperature dependence of the charge imbalance relaxation $Q^{\ast}\sim I_{d}^{ex} \sim \delta\mu_S$ at ultralow temperatures [Fig.~\ref{Arutyunov5}(c)] has a rather simple explanation: at the weak effective electron-phonon interaction the only remaining relaxation channel is the elastic scattering by the inhomogeneity and anisotropy of the energy gap and/ or the finite value of the supercurrent \cite{Tinkham1972}.
In contract to this (transverse) excitation mode the energy relaxation (the longitudinal mode) always requires the inelastic scattering. At ultra-low temperatures $k_BT \ll \Delta$, the concentration of equilibrium phonons is not sufficient for the effective energy relaxation. The only source of phonons with energies larger than the temperature might be the spreading of the Joule heat from the `hot' injector. However, as was already discussed earlier, even in case of maximum injection currents the increase of the electron temperature of the injector is $\delta T_{e}^{i} \simeq$ 100 mK, which  is sufficiently smaller than the minimum energy $2\Delta$ required for the formation of a Cooper pair from an electron-like and a hole-like quasiparticles. Since the energy relaxation is slow, as was found above, but still occurs on `astronomically' large (for a superconductor) scales $\Lambda_{T^{\ast}}\simeq$ 40 $\mu$m, a corresponding scattering channel has to exist. For example, the emission and absorption of nonequilibrium phonons or photons, can provide an appropriate relaxation mechanism. Naturally this relaxation channel also has to make a finite contribution to the charge imbalance relaxation. The observed considerable difference in the characteristic relaxation scales $\Lambda_{Q^{\ast}}<\Lambda_{T^{\ast}}$ confirms our assumption that the elastic scattering is an utterly important relaxation mechanism of the charge imbalance at ultralow temperatures.

A considerable quantitative difference between $\Lambda_{T^{\ast}}$ and $\Lambda_{Q^{\ast}}$ values at ultralow temperatures $k_B T \ll \Delta$ is in the qualitative agreement with previous theoretical calculations \cite{Kaplan1976, Chi1979}. The explicit condition when the elastic scattering can be neglected has been obtained in \cite{Bezuglyi1977},
\begin{equation}
\frac{T_{c}-T}{T_{c}} \ll (\tau_{E}T_{c})^{-2/3},
\end{equation}
where $\tau_E$ is the characteristic time of the energy relaxation (in units $1/K$). This condition allows us to conclude that the only requirement for $\lambda_Q \gg \lambda_E \simeq \lambda_{T^{*}}$ is the high-temperature limit $T \rightarrow T_c$. Formally this observation is equivalent to the statement that the relaxation channel of the charge imbalance does not exist in a normal metal and the corresponding time tends to infinity as the temperature is approaching $T_c$. The experiments discussed above have been conducted at ultralow temperatures $T \ll T_c$, where, to our best knowledge, there are no theoretical predictions to \textit{a priori} relate $\Lambda_{Q^{\ast}}$ with $\Lambda_{T^{\ast}}$.

%%%%%%%%%%%%%%%%%%%%%%%%%%%%%%%%%%%%%%%%%%%%%%%%%%%%%%%%%%%%%%%%%%%%%%%%%%%%
%%%%%%%%%%%%%%%%%%%%%%%%%%%%%%%%%%%%%%%%%%%%%%%%%%%%%%%%%%%%%%%%%%%%%%%%%%%%

\section{Nonlocal supercurrent measurements for charge imbalance length determination}\label{nonlocal}

%%%%%%%%%%%%%%%%%%%%%%%%%%%%%%%%%%%%%%%%%%%%%%%%%%%%%%%%%%%%%%%%%%%%%%%%%%%%
%%%%%%%%%%%%%%%%%%%%%%%%%%%%%%%%%%%%%%%%%%%%%%%%%%%%%%%%%%%%%%%%%%%%%%%%%%%%

%

\begin{figure*}[t!]\label{StolyarovV1}
\begin{center}
\epsfxsize=16cm\epsffile{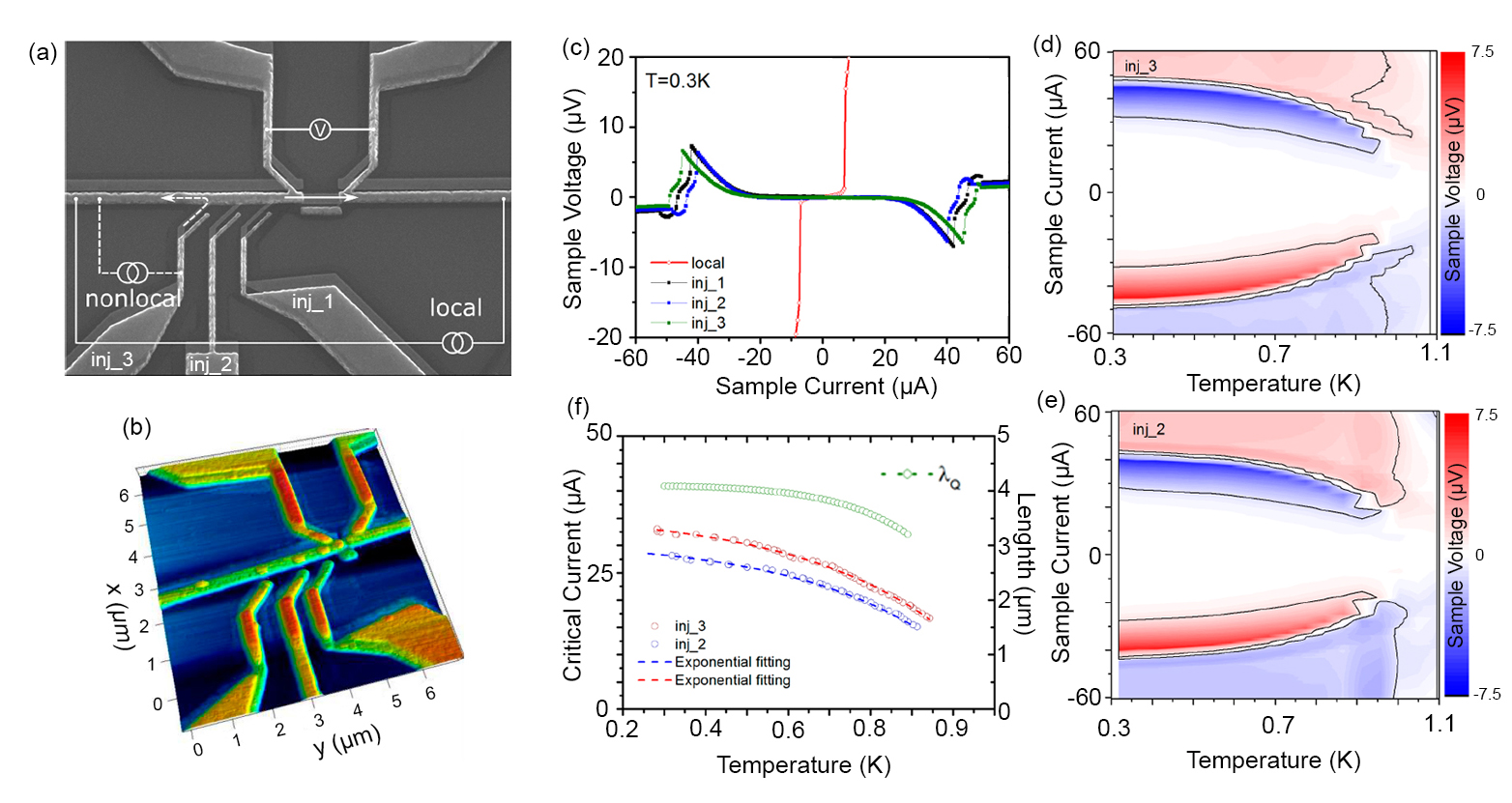}
\end{center}
\caption{(a) SEM image of planar Al/Cu/Al Josephson junction with a length of copper section 700 nm and a measurement scheme with three SN injectors; (b) 3D image plotted by using Atomic Force Microcopy; (c) The typical IV curves for local and nonlocal type of measurements depicted in (a) structure; (d,e) Evolution of IV curves with temperature for two different injectors correspondingly; (f) Temperature dependence of charge imbalance relaxation length for Al superconducting leads.}
\end{figure*}

In this section we discuss another method of the experimental determination of charge-imbalance relaxation length. It can be estimated using a modified Kadin, Smith and Skocpol (KSS) scheme \cite{Kadin1980} in case of a planar geometry. We refer to recent works, done in the group of Ryazanov \cite{Golikova2014}. In this paper the detection of a nonlocal critical current was demonstrated in mesoscopic SNS (Al-Cu-Al) Josephson junctions with several spatially separated normal metal (Cu) injectors connected to one of the superconducting Al wires. Similar structure is shown in Fig.~\ref{StolyarovV1}(a). To measure the nonlocal voltage at low temperatures $T \ll T_c$, we need to use superconducting leads just near the junction (wite lines in Fig.~\ref{StolyarovV1}(a)).  In order to describe the interplay of charge imbalance and Josephson effect in the realized mesoscopic system in the low-temperature limit, a two-channel charge-imbalance KSS model was proposed \cite{Kadin1980}. This approach was simplified for the case of low-frequency processes and extended to study the effect of nonequilibrium quasiparticle flow in Josephson SNS junctions by Kaplunenko, Ryazanov, and Schmidt \cite{Kapulenko1985}. The authors assume  that nonequilibrium processes in a superconductor including the conversion of a quasiparticle flow into a pair current can be described reasonably by means of an equivalent circuit introduced by Kadin, Smith, and Skocpol for the explanation of phase-slip-center behavior. Recently, in \cite{Golikova2014} this model was modified for the planar Josephson structures geometry.

To determine the charge imbalance length in the multi-terminal planar structure, Golikova et al. derived a simple equation \cite{Golikova2014},
\begin{equation}
\Lambda_{Q^{\ast}}=\frac{d_{inj_3}-d_{inj_2}}{\ln(I_c^{inj_3}/I_c^{inj_2})},
\end{equation}
where $\Lambda_{Q^{\ast}}$ is the charge-imbalance relaxation length, $d_{inj_3}$ and $d_{inj_2}$ are the distances between planar Josephson junction and injector probes, correspondingly (Fig.~\ref{StolyarovV1}(a) white lines), $I_c^{inj_3}$ and $I_c^{inj_2}$ are the critical currents of planar Josephson junction in case of nonlocal measurements scheme (Fig.~\ref{StolyarovV1}(c,d,e,f)).

The results presented in Fig.~\ref{StolyarovV1} were recently obtained in the group of Stolyarov by a measurement technique similar to that of \cite{Golikova2014}. Fig.~\ref{StolyarovV1}(a) shows a scanning electron microscopy (SEM) image of our sample, together with a measurement scheme. The submicron planar structures were fabricated by means of electron beam lithography and standard \emph{in situ} shadow e-beam evaporation without breaking the vacuum. In our case the copper layer thickness was $d_{Cu}$ = 30 nm and the aluminum thickness $d_{Al}$ = 250 nm, correspondingly. All injectors were realized as long copper SNS junctions without Josephson coupling. Geometrical size of SNS elements can be seen in Atomic Force Microscope (AFM) image, see Fig.~\ref{StolyarovV1}(b).

Figure~\ref{StolyarovV1}(c) depict the current-voltage (IV) dependences for local (red curve) and nonlocal scheme of measurements for three cases of injector position (blue, black and green yellow, correspondingly). Figure~\ref{StolyarovV1}(d,e) presents an evolution of IV curves with temperature for two different injectors (inj$_2$ and inj$_3$). The critical current dependance in nonlocal measurements for $I_c^{inj_3}$ and $I_c^{inj_2}$ allows as to obtain the charge imbalance relaxation length for Al superconducting wires. In our case it varies in the range from 3~$\mu$m to 4~$\mu$m in temperature range from 0.2~K to 0.9~K. Such transport measurements are simple enough for determination of charge imbalance relaxation length in differen type superconductors. The obtained value of $\Lambda_{Q^{\ast}} \sim 4\mu$m is consistent with results, obtained in the group of Arutyunov (see section \ref{imbalance} \cite{Arutyunov}).

%%%%%%%%%%%%%%%%%%%%%%%%%%%%%%%%%%%%%%%%%%%%%%%%%%%%%%%%%%%%%%%%%%%%%%%%%%%%
%%%%%%%%%%%%%%%%%%%%%%%%%%%%%%%%%%%%%%%%%%%%%%%%%%%%%%%%%%%%%%%%%%%%%%%%%%%%

\section{Nonequilibrium electron cooling}\label{cooling}

%%%%%%%%%%%%%%%%%%%%%%%%%%%%%%%%%%%%%%%%%%%%%%%%%%%%%%%%%%%%%%%%%%%%%%%%%%%%
%%%%%%%%%%%%%%%%%%%%%%%%%%%%%%%%%%%%%%%%%%%%%%%%%%%%%%%%%%%%%%%%%%%%%%%%%%%%

%

The subject of nonequilibrium quasiparticle relaxation is of primary importance
for the operation of normal metal - insulator - superconductor (NIS) refrigerators (see the review papers \cite{Muhonen2012, Giazotto2006, Virtanen2007, Courtois2014} and references therein).
In such devices the flow of the electric current is accompanied by heat transfer from the normal metal into the superconductor,
which enables cooling of electrons in the normal metal (see section \ref{f_noneq}).
This phenomenon arises due to selective tunneling of high-energy quasiparticles out of the normal
metal which is induced by the superconducting energy gap.
Though the principle of operation is rather straightforward, the high cooling power requires
high density of nonequilibrium quasiparticles injected into the superconductor
and accumulated near the tunnel interface \cite{Vasenko2009, Rajauria2012}. The
consequences are the backtunneling of hot quasiparticles to
the normal metal \cite{Vasenko2009, Jug2001}, the emission of phonons that partially penetrate the normal metal \cite{Rajauria2012, Jug2001},
and the overheating of the superconducting electrode \cite{Rajauria2012}. All these effects reduce
the efficiency of NIS refrigerators. To minimize the undesired
back-action, relaxation of those nonequilibrium quasiparticles should happen in the
superconductor electrode \cite{Vasenko2009, Rajauria2012}.

A micrometer-sized refrigerator, based on a NIS tunnel junction, has been first fabricated by Nahum et al. \cite{Nahum1994}. The authors
used a single NIS junction in order to cool a small normal metal strip. Later, Leivo et al. \cite{Leivo1996} noticed that the cooling
power is an even function of the applied voltage and fabricated a refrigerator with two NIS junctions arranged in a
symmetric series configuration (SINIS). Another important point is that in SINIS geometry the normal metal lead is effectively isolated
by two superconductor electrodes from the electromagnetic environment which improve cooling performance.

The quasiparticle poisoning of SINIS refrigerators can be eliminated by a special trick
- utilization of a quasiparticle drain (so called quasiparticle trap),
and by fine tuning the parameters of the NIS junctions tunnel barrier. The drain should be `isolated' from the
superconductor with thin tunnel barrier, which from one side lets quasiparticles get efficiently trapped
in the drain, while from another side, stops the inverse proximity effect. In particular, it has
been demonstrated that  electronic cooling can be optimized in specially-designed large area
normal metal-insulator-superconductor junctions \cite{Courtois2013}.
The two key ingredients were found to be of high importance: (i) the tunnel barrier transparency
for the cooling junctions \cite{Nguyen2013},
(ii) the coupling to a quasiparticle drain, through a (separate) tunnel junction \cite{Kauppila2013}.
With such provisions  temperature reduction of a factor 5, from 150 mK down to 30 mK,
and a cooling power of the order of one nanowatt has been achieved. An additional improvement
of cooling performance can be reached by utilization of the second-stage SINIS cooler actively
evacuating quasiparticles out of the hot superconductor, especially in the low-temperature limit \cite{Nguyen2016}.
The working principle of the device is following: back sides of the main cooler are connected to two other SINIS coolers.
These second-stage coolers act as ``active quasiparticle traps'' and help to thermalize the hot superconductor leads of the main device.

Rather unusual result has been recently obtained in electron refrigerators utilizing
semiconductor - insulator - superconductor junctions \cite{Gunnarsson2015}.
Such an interface barrier does not increase the junction resistance
but strongly reduces the detrimental sub-gap leakage current. The result was attributed
to the Fermi level de-pinning and dopant segregation effects that strongly affect the
junction properties at the nanoscale. Due to high transparency and low leakage such
semiconductor - insulator - superconductor junctions showed excellent
cooling power performance, comparable to that of high power NIS coolers.

%%%%%%%%%%%%%%%%%%%%%%%%%%%%%%%%%%%%%%%%%%%%%%%%%%%%%%%%%%%%%%%%%%%%%%%%%%%%
%%%%%%%%%%%%%%%%%%%%%%%%%%%%%%%%%%%%%%%%%%%%%%%%%%%%%%%%%%%%%%%%%%%%%%%%%%%%

\section{Summary and outlook}\label{summary}

%%%%%%%%%%%%%%%%%%%%%%%%%%%%%%%%%%%%%%%%%%%%%%%%%%%%%%%%%%%%%%%%%%%%%%%%%%%%
%%%%%%%%%%%%%%%%%%%%%%%%%%%%%%%%%%%%%%%%%%%%%%%%%%%%%%%%%%%%%%%%%%%%%%%%%%%%

In conclusion, recently there have been performed spatially resolved measurements of the
nonequilibrium quasiparticle relaxation in superconducting aluminum. For the first time on
the same hybrid microstructures made of aluminum and copper the spatial, temperature and energy
characteristics of the energy and charge imbalance were measured at ultralow temperatures, $T\ll T_{c}$.
This imbalance occurs when nonequilibrium quasiparticle excitations are injected into the superconductor
form a normal metal through the tunnel barrier.

It was shown that the experimental results can be described by the phenomenological model which assumes
the validity of the equilibrium expression for the NIS junction tunnel current. Further, it has been postulated
equilibrium functional dependencies of the density of states (DOS) and distribution function of the superconductor,
while assumed that the DOS broadening parameter (Dynes parameter) $\Gamma (I_i, L_{id}, T)$, the effective chemical potential of the Cooper pairs
$\mu_S (I_i, L_{id}, T)$, and the superconducting gap $\Delta (I_i, L_{id}, T)$ depend on the rate of quasiparticle injection
$I_i$ and the distance to the injector junction $L_{id}$ for a given temperature $T$.

It was also shown that spatial relaxations of the nonequilibrium quasiparticle excitations in the aluminum can be described
by the exponential dependence with a typical scale of $\Lambda_{T^{\ast}}=$ $40\pm20$ $\mu$m and $\Lambda_{Q^{\ast}}$ $=5\pm1.5$ $\mu$m
for the energy (longitudinal) and charge (transverse) imbalance, respectively.
It should be stressed that both quantities were measured simultaneously on the same samples and
using the same experimental technique, which eliminates various sample and measurement artifacts.

This is the central result of this review. It should be emphasized that both the energy and the charge disequilibrium modes are universal phenomena
which should be taken into consideration in a broad class of systems involving electron, spin and/ or coherent non-local transport. Despite the
reasonable agreement with the phenomenological model, application of the relaxation approximation approach for the essentially spatially inhomogeneous problem is not fully justified at ultra-low temperatures, $T\ll T_{c}$. A deeper (microscopic) model is required for such a quantitative analysis. We hope that the findings presented in this review will trigger the corresponding research activity.

This result allows us formulate an interesting hypothesis. We have found that in superconducting Aluminum both transversal and longitudinal modes relax on `astronomical' scale for superconductors, where the coherence length 100nm gives the characteristic scale. Based on this one can make a hypothesis that at some conditions one can detect the coherent transport of nonequilibrium quasiparticles. It can be studied, for example, in Aaronov-Bohm type experiment.

We have also presented new data on nonlocal measurements of charge imbalance relaxation length. The results are consistent with \cite{Arutyunov}, with $\Lambda_{Q^{\ast}} \sim 4\mu$m. Such large scales of charge and energy imbalance relaxation impose important limits for high-density integration of logical superconducting circuits and should be taken into account. In this paper we have also reviewed recent articles on nonequilibrium electron cooling in NIS refrigerators. We discussed the problems of quasiparticle poisoning and quasiparticle evacuation by so called quasiparticle traps.

\ack{The authors would like to thank F S Bergeret, V V Ryazanov and A D Zaikin for fruitful discussions. The experimental part of paper has been reviewed by KYuA, who acknowledges the support of Russian Science Foundation Grant No. 16-12-10521 ``Quantum fluctuations in superconducting nanostructures''. The theory part review has been prepared by ASV and two students, SAC and TIK, who acknowledge the support of joint Russian-Greek projects RFMEFI61717X0001 and T4$\Delta$P$\Omega$-00031 ``Experimental and theoretical studies of physical properties of low dimensional quantum nanoelectronic systems''. Section 11 was reviewed by DSL and VSS. VSS acknowledges RFBR Project No. 16-02-00815 A. ASV acknowledge the hospitality of Donostia International Physics Center (DIPC) during his stay in Spain.}

\end{document}